\documentclass[fleqn,usenatbib]{mnras}

\usepackage[T1]{fontenc}

\usepackage{graphicx}	
\usepackage[normalem]{ulem} 
\usepackage{scalefnt}

\usepackage{amsmath}	
\usepackage{amssymb}	
\usepackage{multirow}

\defcitealias{Galan2023}{Paper~I}

\newcommand{\soutPC}{\bgroup\markoverwith{\textcolor{cyan}{\rule[0.5ex]{2pt}{1pt}}}\ULon}

\usepackage{adjustbox}
\usepackage{subfig}

\newcommand{\placetabMergersNEW}{
\begin{table}
\caption{Main parameters of the mergers.}     
\label{tab:merger_props}
\centering
\begin{tabular}{c c c c}     
\hline
Galaxy merger\;\;\;\;\; & $R_{\rm initial}$ [kpc]\;\;\;\;\; & $r^{(\rm first)}_{\rm peri}$ [kpc]\;\;\;\;\; & $i$ [deg] \\
\hline \\ \vspace{1mm}
   co-co & \multirow{5}{1cm}{$378.8$} & $36.5$ & 0 \\
   45-tilted & & $37.1$ & 45 \\
   polar & & $36.5$ & 90 \\
   135-tilted & & $36.8$ & 135 \\
   co-ret & & $37.0$ & 180 \\
\hline    
\end{tabular}
\begin{minipage}{8.4cm}
{\em Notes.} Col. (1) galaxy merger name. Col. (2) initial separation between the galaxies. Col. (3) first pericentric passage. Col. (4) inclination angle of the orbit.
\end{minipage}
\end{table}
}

\newcommand{\placetabPropertiesNEW}{
\begin{table}
\caption{Properties of the merging galaxies.}     
\label{tab:secondary_props}
\centering
\begin{tabular}{c c c c c c}     
\hline
Galaxy & $M_{\rm total}$ & $R_{\rm cutoff}$ & $M_{\rm BH}$ & $M_{\star}$ & $R_{\rm 50}$ \\
 & [$10^{11}$~M$_{\sun}$] & [kpc] & [$10^{6}$~M$_{\sun}$] & [$10^{9}$~M$_{\sun}$] & [kpc] \\
\hline \\ \vspace{1mm}
   Primary & $24.8$ & $243.0$ & $31.1$ & $27.2$ & $1.68$ \\
   Secondary & $6.2$ & $135.8$ & $7.8$ & $6.8$ & $1.25$ \\
\hline    
\end{tabular}
\begin{minipage}{8.4cm}
{\em Notes.} Col. (1) galaxy name. Col. (2) total mass. Col. (3) cutoff radius of the DM halo. Col. (4) mass of the central SMBH. Col. (5) total stellar mass. Col. (6) half-mass radius of the stellar component.
\end{minipage}
\end{table}
}

\newcommand{\placefigKinem}{
\begin{figure*}
    \centering
    \includegraphics[scale=0.67]{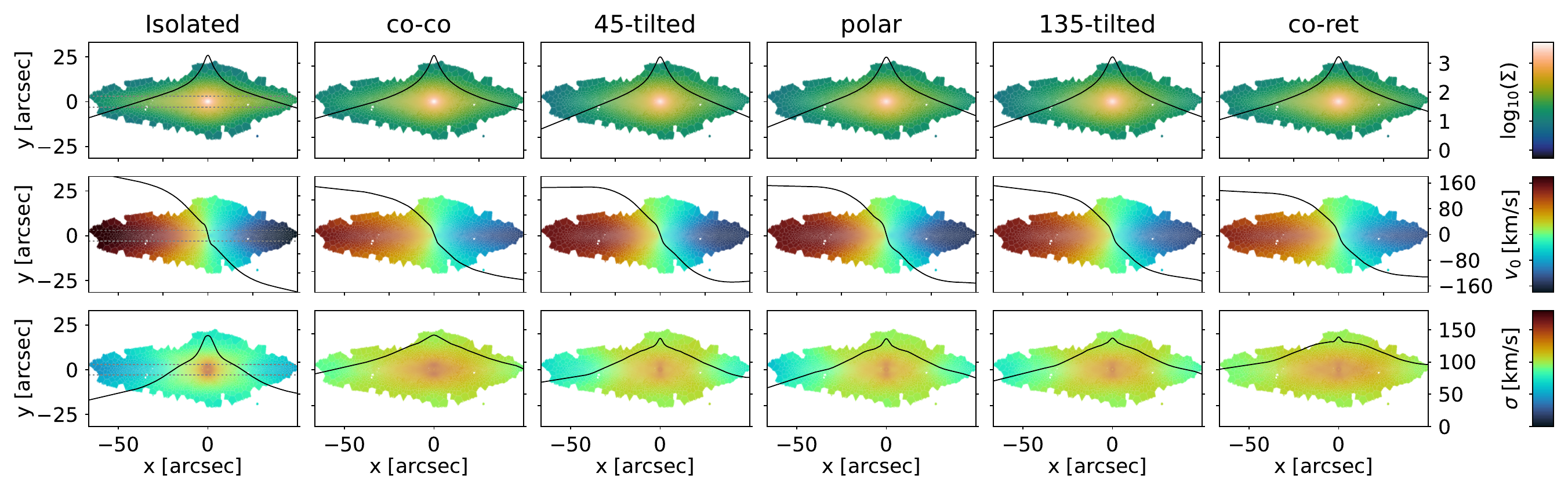}
    \caption{Comparison of the FCC\,170 $N$-body model evolved in isolation at 10~Gyr (left-hand column) with the merger remnants (second to sixth columns), as specified on top of each column. From top to bottom, the rows show the maps of the surface stellar luminosity density (top panels), mean velocity (middle panels), and velocity dispersion (bottom panels) of the Gauss--Hermite moments. The solid lines correspond to the radial profiles extracted along the galaxy major axis in the region bracketed by the horizontal dotted lines. At the distance of the Fornax cluster \citep[21.9~Mpc;][]{Pinna2019}, 1~kpc corresponds to 9.4~arcsec.
    }
    \label{fig:kinematics_mergers}
\end{figure*}
}

\newcommand{\placefigDistances}{
\begin{figure}
    \centering
    \includegraphics[scale=0.36]{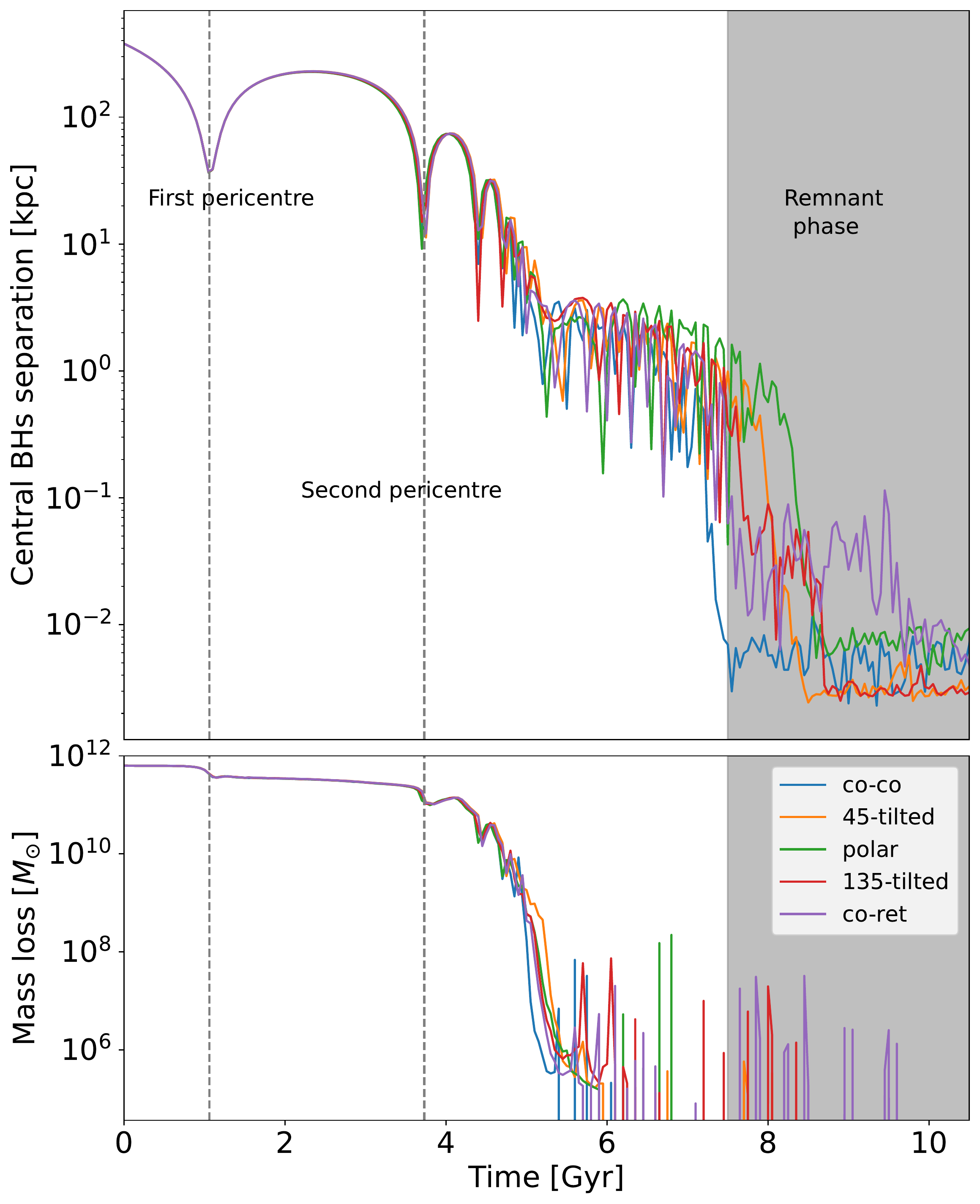}
    \caption{Top panel: evolution of the SMBHs separation for all our mergers. Bottom panel: total mass loss during the mergers. The mass of the secondary SMBH is not included in the calculation of the mass loss. The beginning of the remnant phase is defined as the first time when the SMBH separation is below 100~pc. This varies amongst mergers: for example, the remnant phase occurs sooner for the co-co merger than for the polar merger.
    }
    \label{fig:distance}
\end{figure}
}

\newcommand{\placefigMerger}{
\begin{figure*}
    \centering
    \includegraphics[scale=0.45]{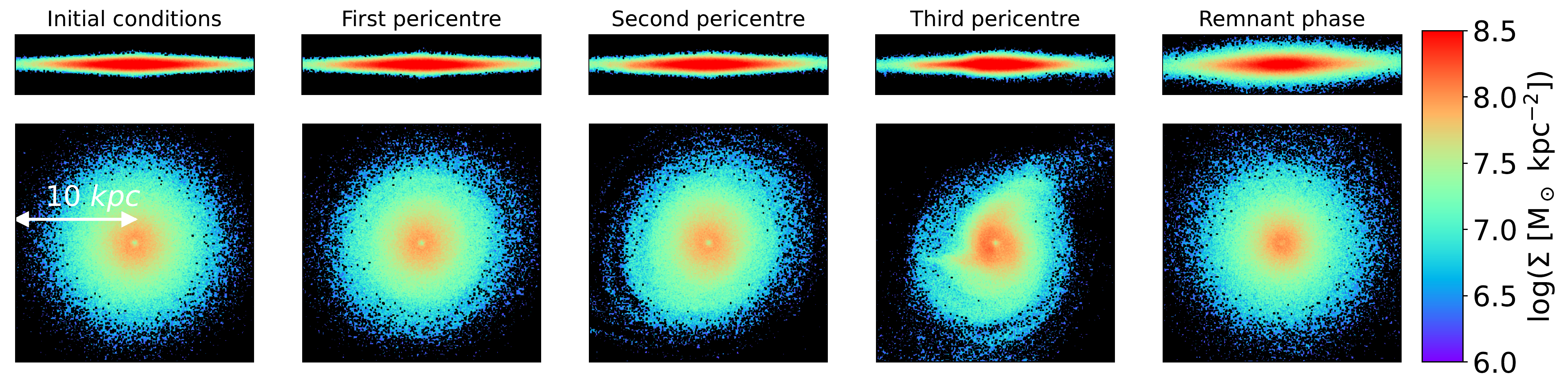}
    \caption{Edge-on view (top panels) and face-on view (bottom panels) of the surface mass density of the thin-disc particles of the primary galaxy for the 1:4 co-planar co-rotating encounter, at different stages. The remnant phase represents the resulting thin disc after the two central SMBHs sink. Each panel is centred on the primary central SMBH.
    }
    \label{fig:merger}
\end{figure*}
}

\newcommand{\placefigMergerNSD}{
\begin{figure*}
    \centering
    \includegraphics[scale=0.45]{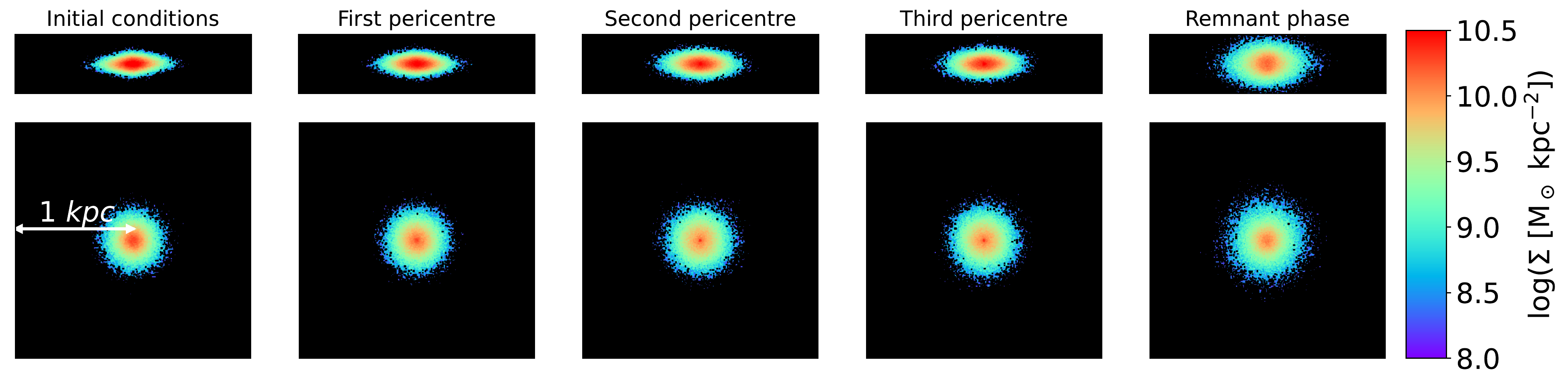}
    \caption{Same as Figure~\ref{fig:merger}, but for the NSD particles of the primary galaxy.
    }
    \label{fig:merger_NSD}
\end{figure*}
}

\newcommand{\placefigVoronoiCo}{
\begin{figure*}
    \centering
    \subfloat{\includegraphics[scale=0.34]{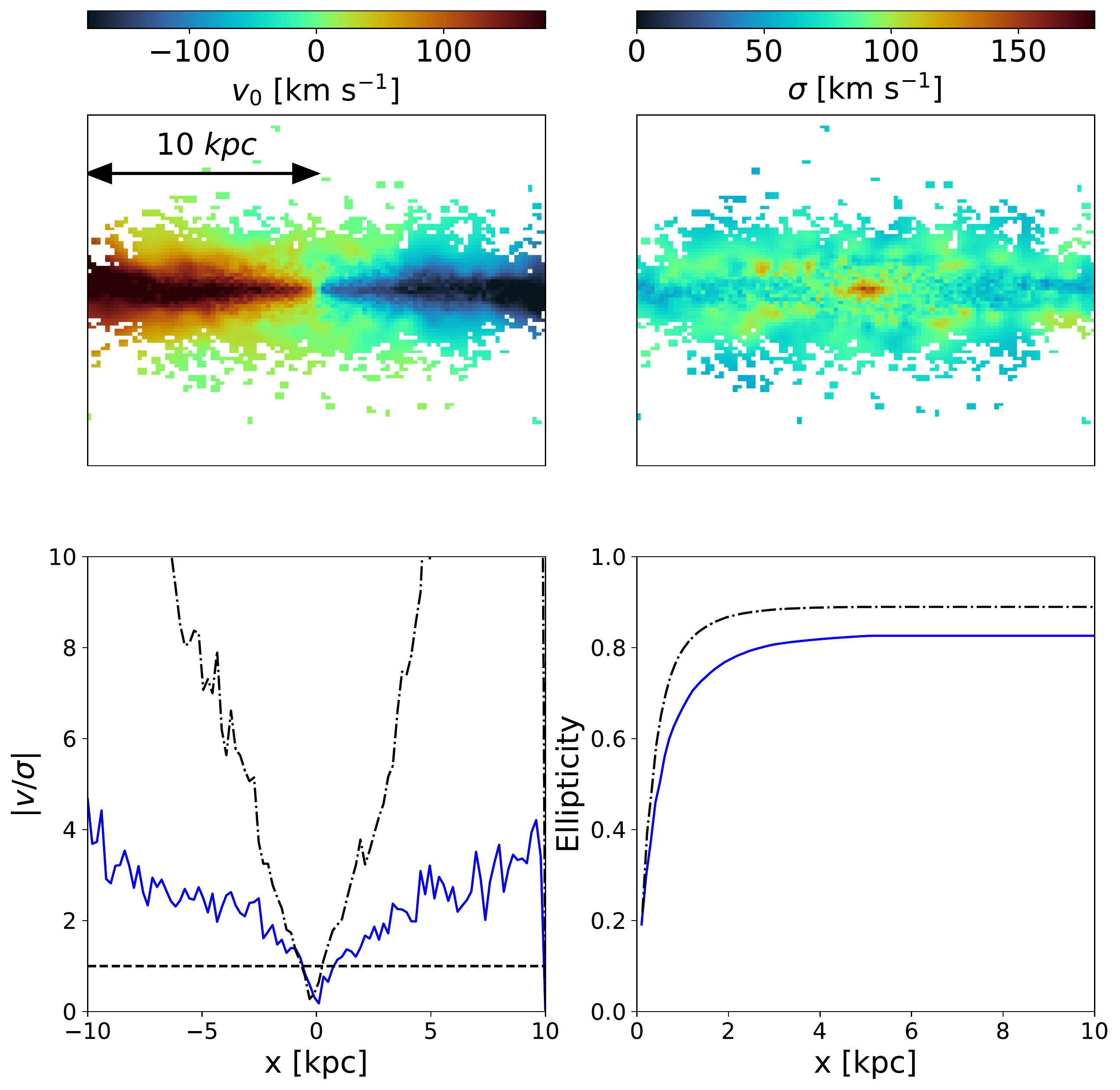}}
    \subfloat{\includegraphics[scale=0.34]{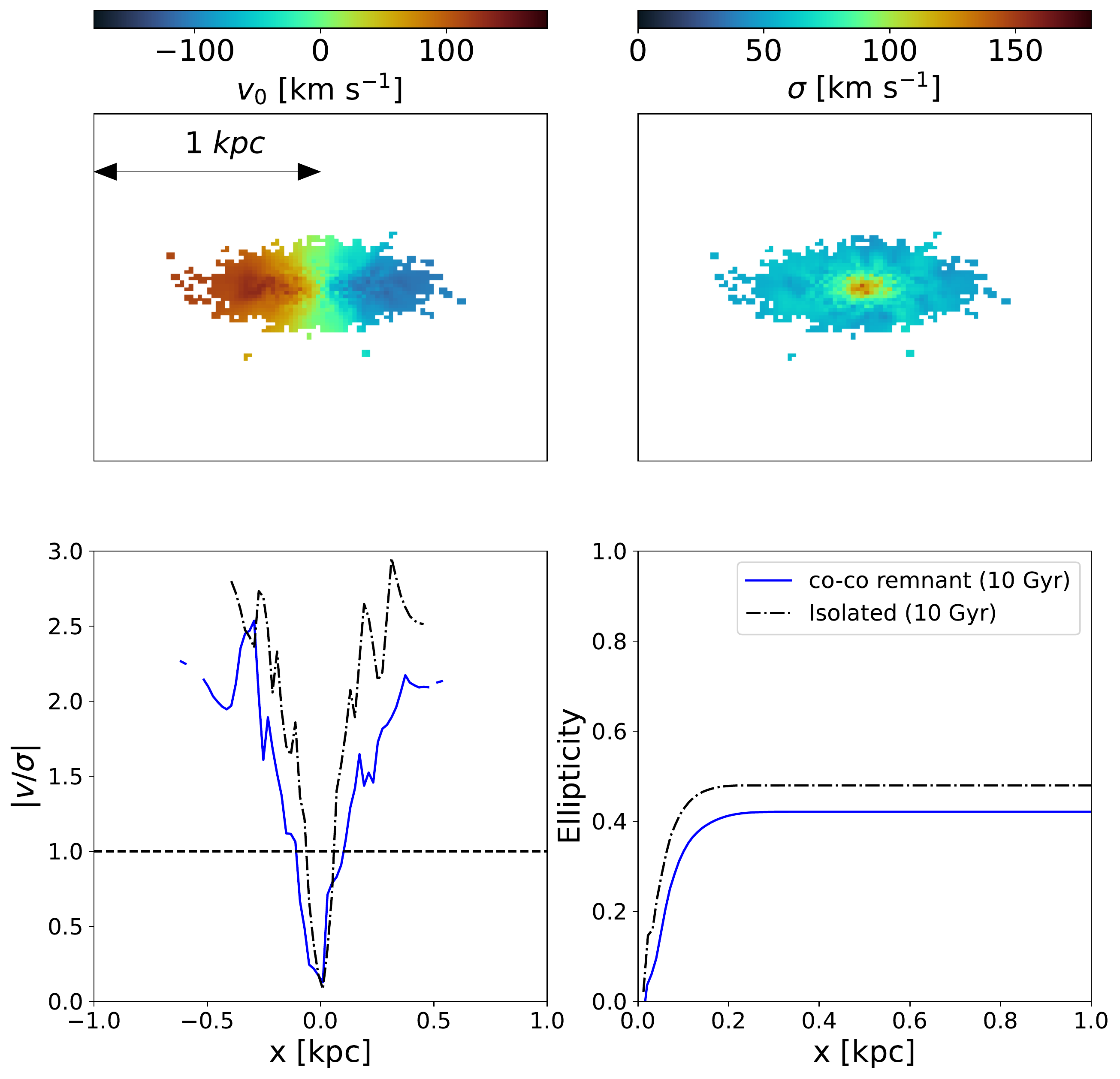}}
    \caption{Top panels: 2D mock images of the mean velocity and velocity dispersion for the thin disc (left-hand panels) and NSD (right-hand panels), both in edge-on view, for the remnant phase (at 10~Gyr) of the co-co merger. Bottom panels: mean velocity over velocity dispersion along the equatorial plane and ellipticity for the thin disc and NSD of the merger remnant (blue solid lines) and FCC\,170 $N$-body model in isolation at 10~Gyr (black dotted-dashed lines).
    }
    \label{fig:voronoi_co-co}
\end{figure*}
}

\newcommand{\placefigVoronoiTiltedCo}{
\begin{figure*}
    \centering
    \subfloat{\includegraphics[scale=0.34]{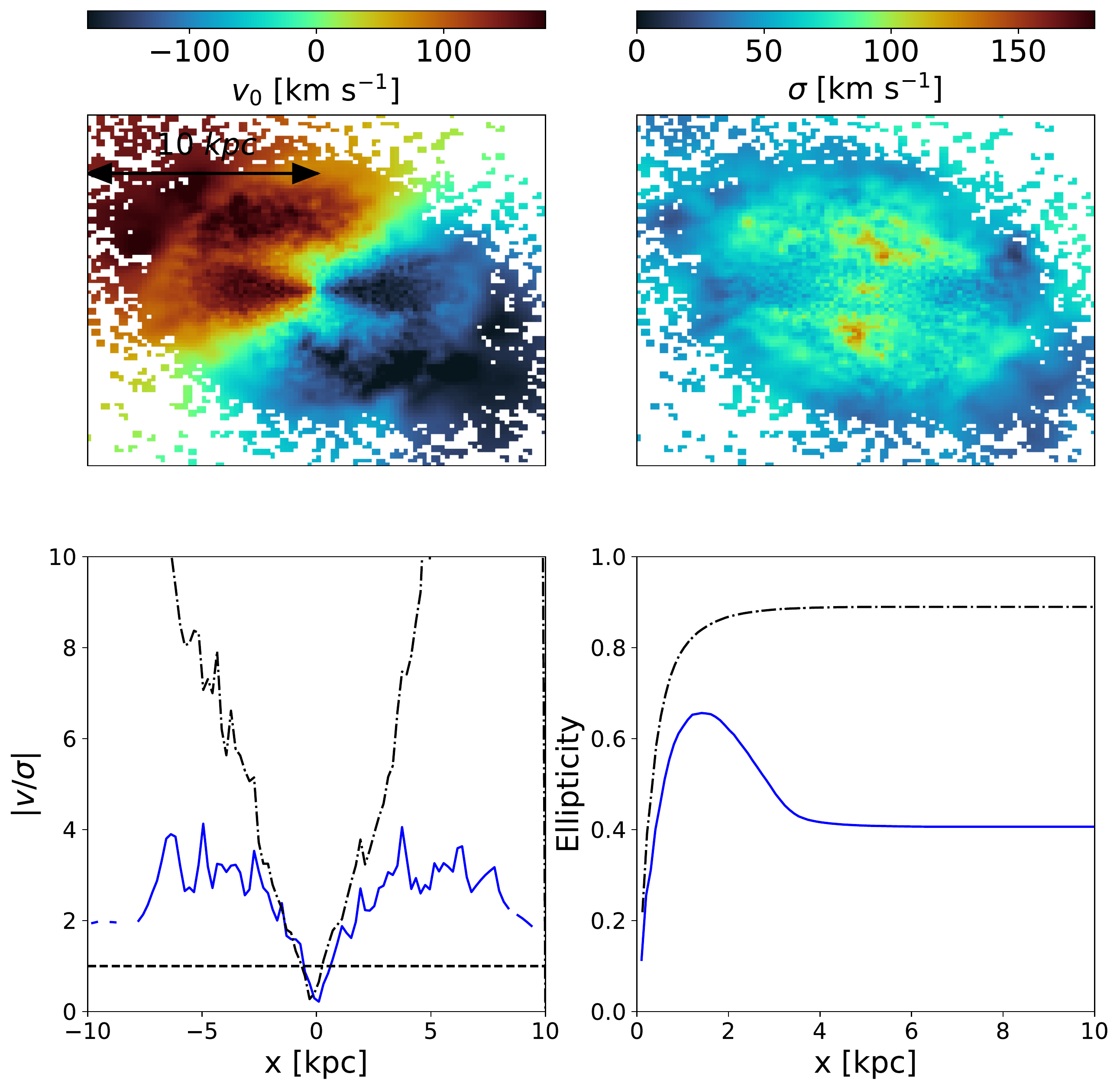}}
    \subfloat{\includegraphics[scale=0.34]{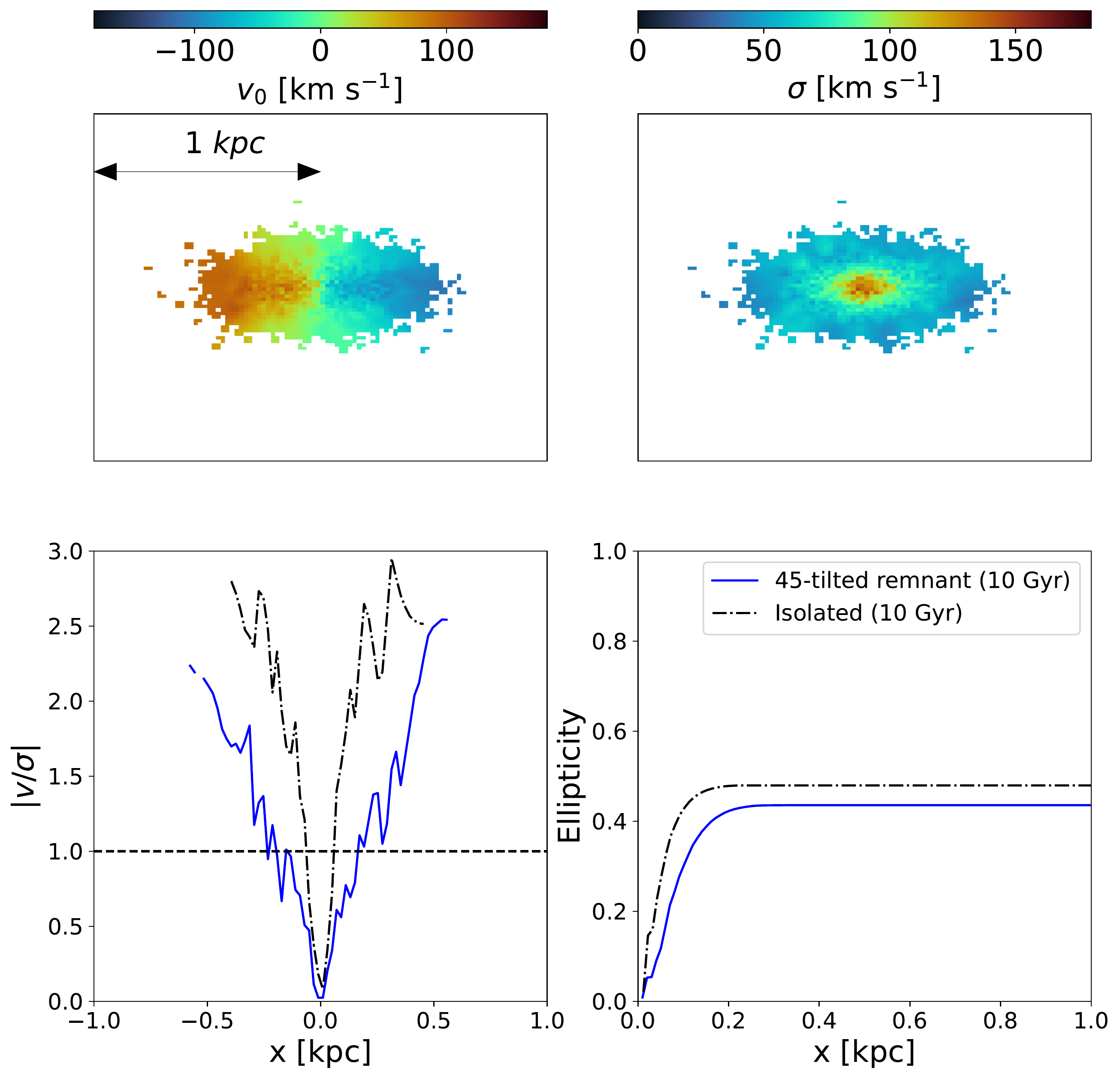}}
    \caption{Same as Figure~\ref{fig:voronoi_co-co}, but for the 45-tilted merger.}
    \label{fig:voronoi_tilted-co}
\end{figure*}
}

\newcommand{\placefigVoronoiPolar}{
\begin{figure*}
    \centering
    \subfloat{\includegraphics[scale=0.34]{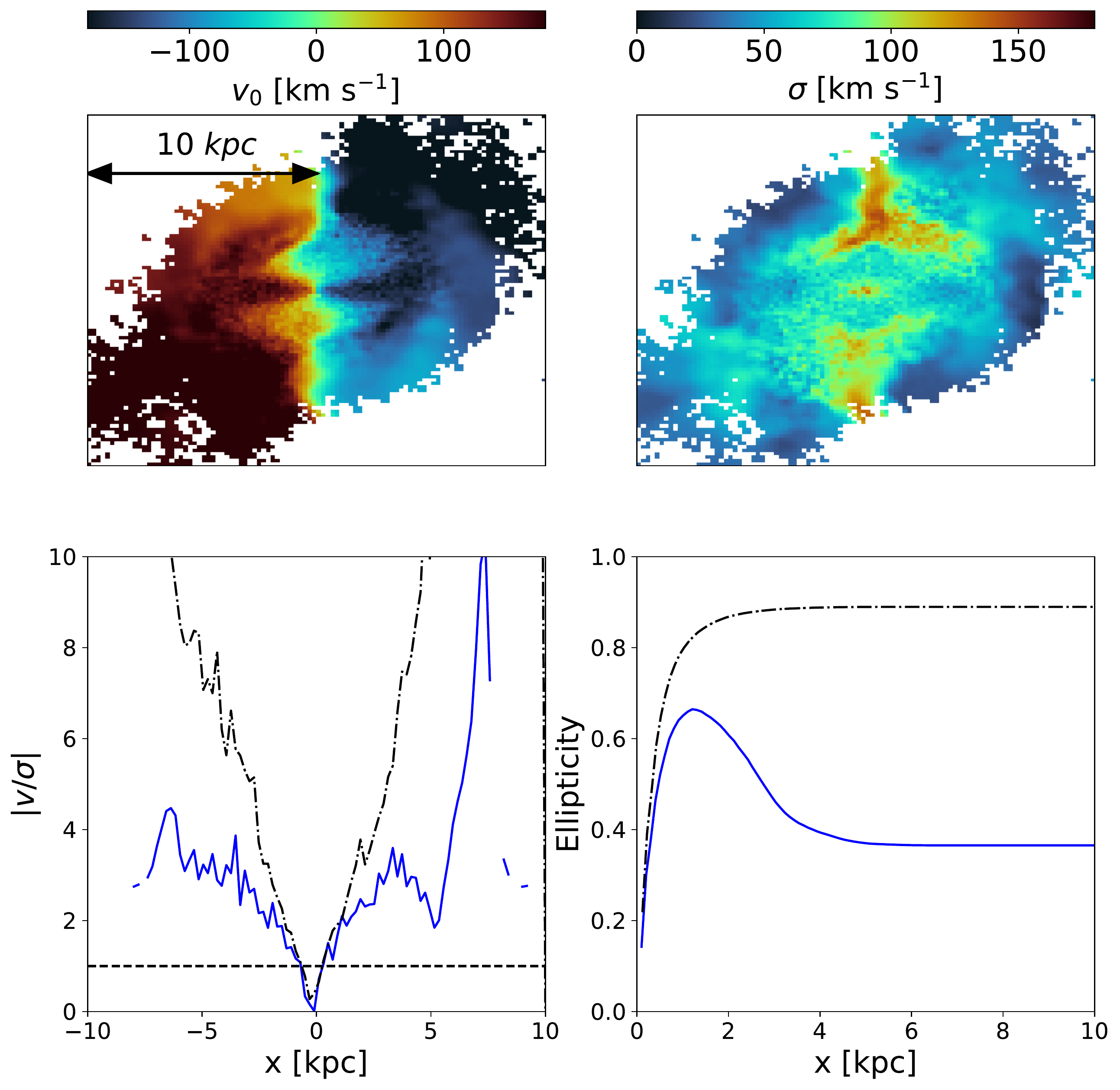}}
    \subfloat{\includegraphics[scale=0.34]{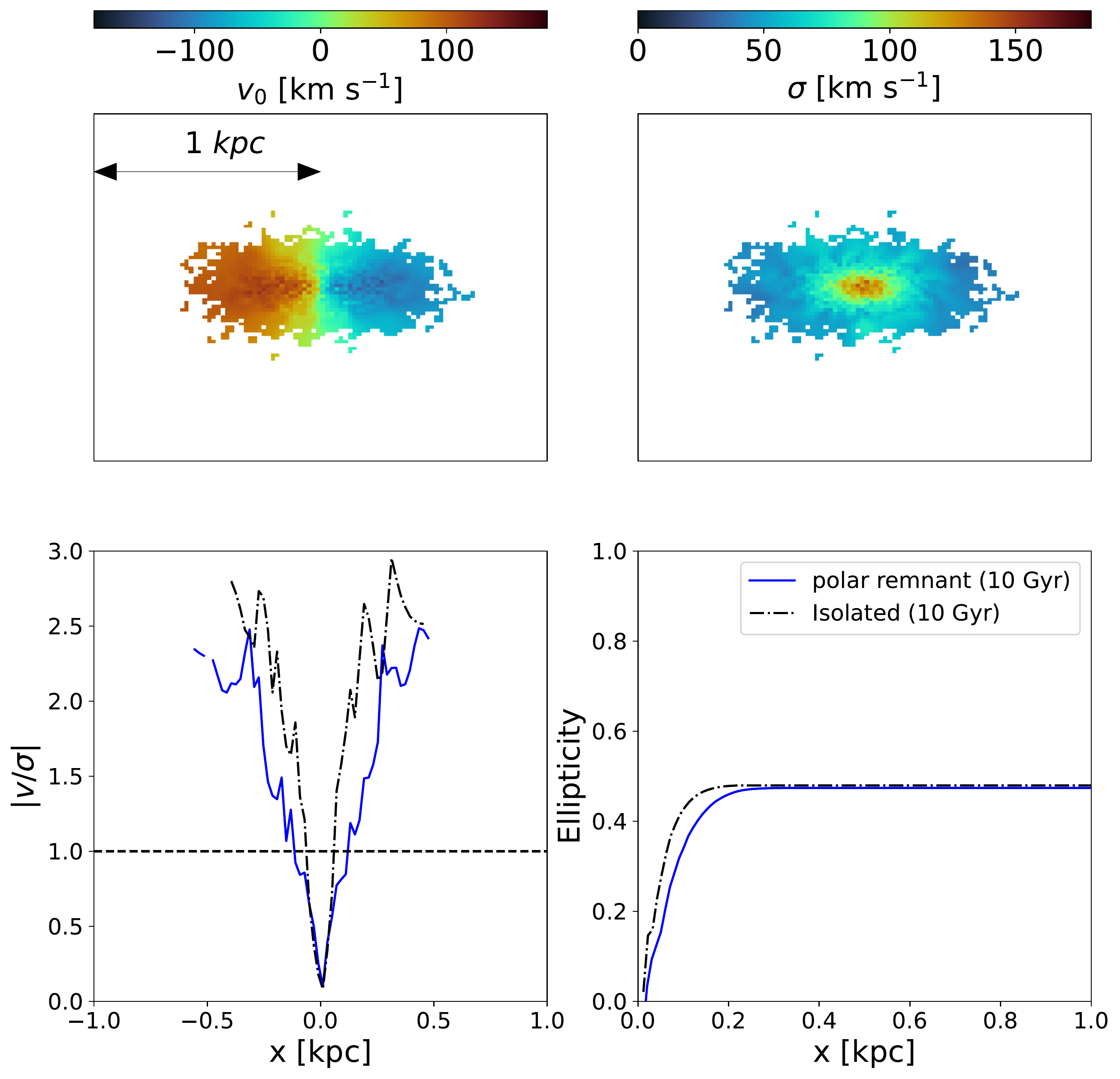}}
    \caption{Same as Figure~\ref{fig:voronoi_co-co}, but for the polar merger.}
    \label{fig:voronoi_polar}
\end{figure*}
}

\newcommand{\placefigVoronoiTiltedRet}{
\begin{figure*}
    \centering
    \subfloat{\includegraphics[scale=0.34]{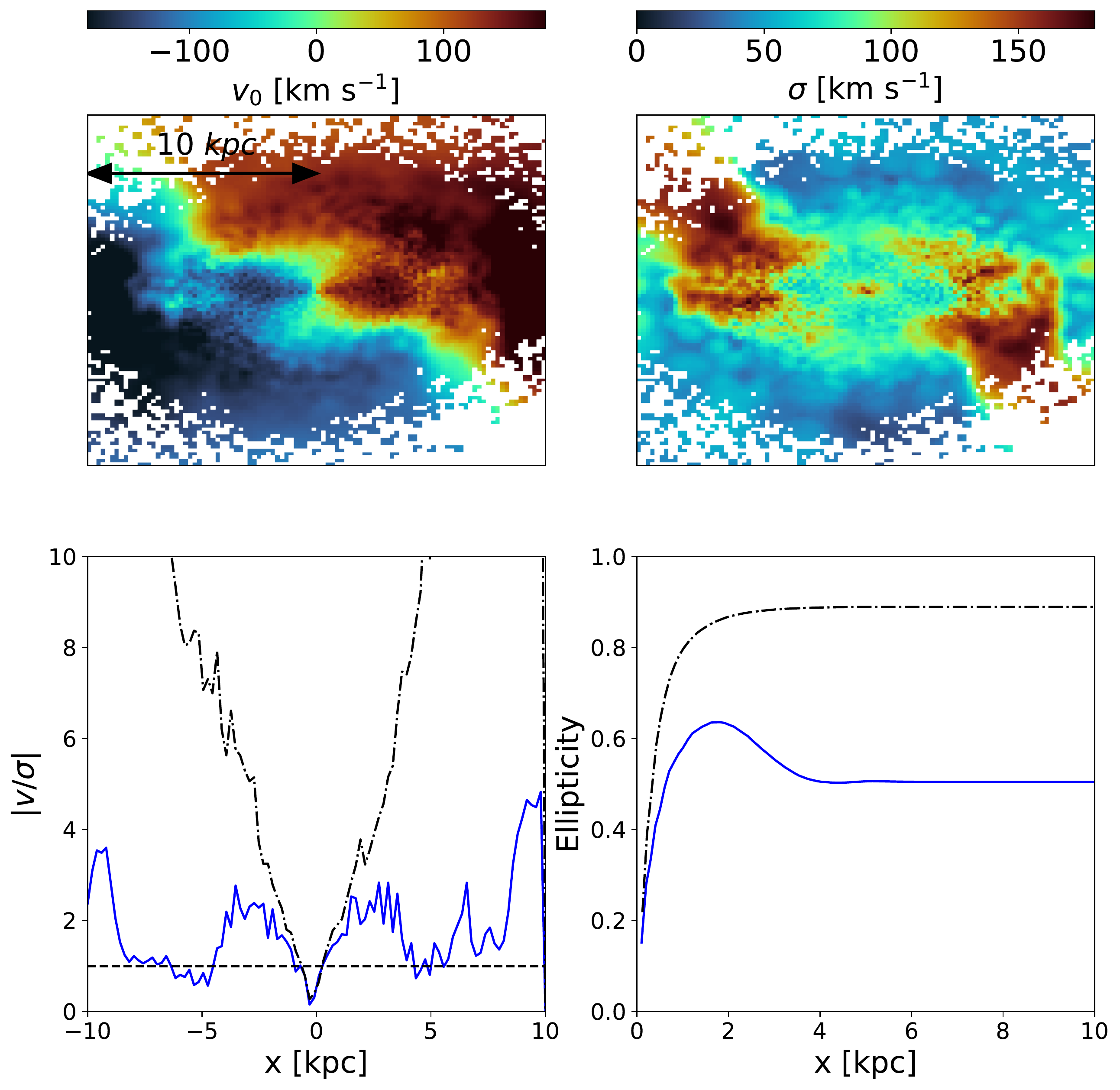}}
    \subfloat{\includegraphics[scale=0.34]{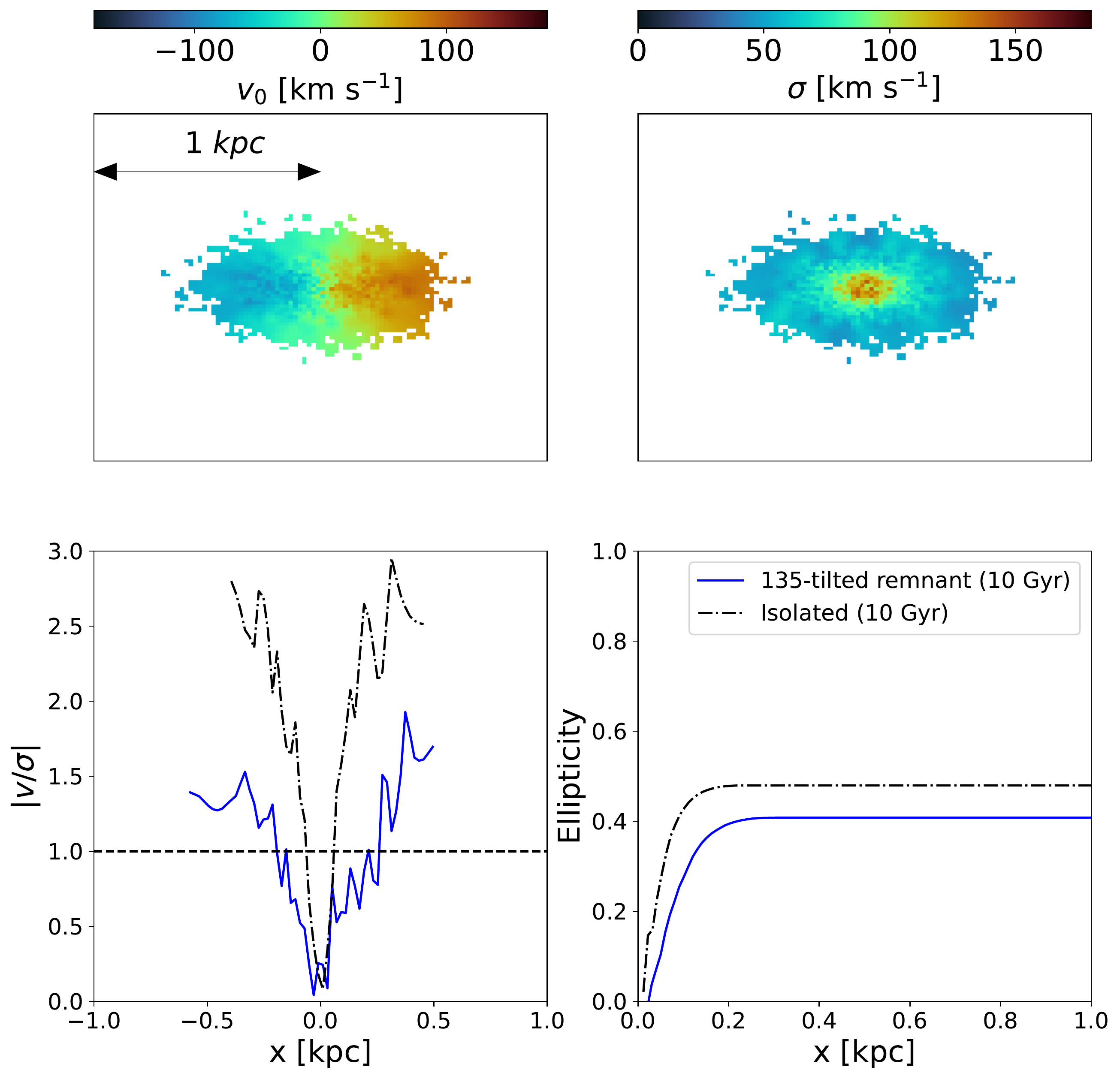}}
    \caption{Same as Figure~\ref{fig:voronoi_co-co}, but for the 135-tilted merger.}
    \label{fig:voronoi_tiled-ret}
\end{figure*}
}

\newcommand{\placefigVoronoiRet}{
\begin{figure*}
    \centering
    \subfloat{\includegraphics[scale=0.34]{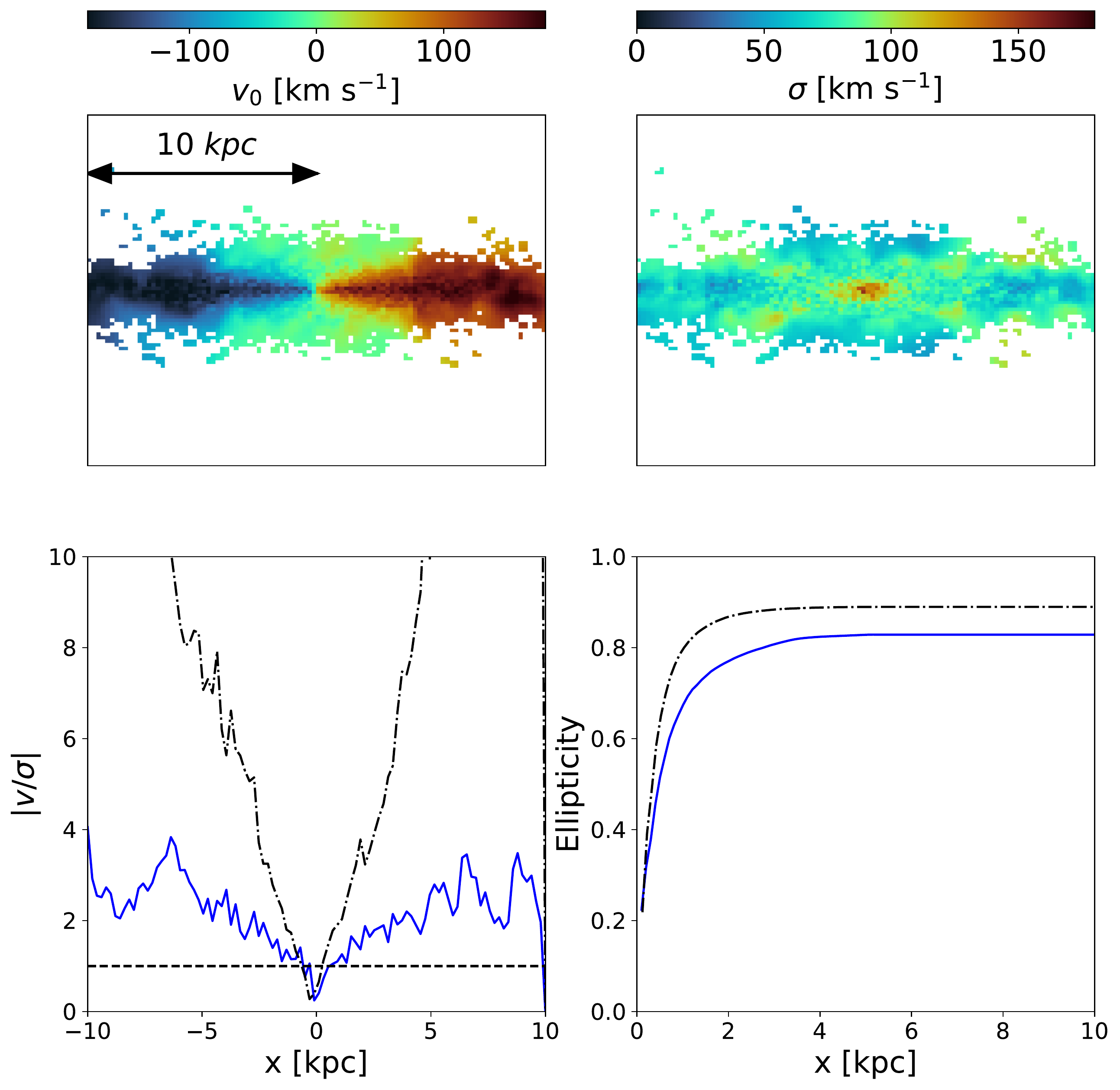}}
    \subfloat{\includegraphics[scale=0.34]{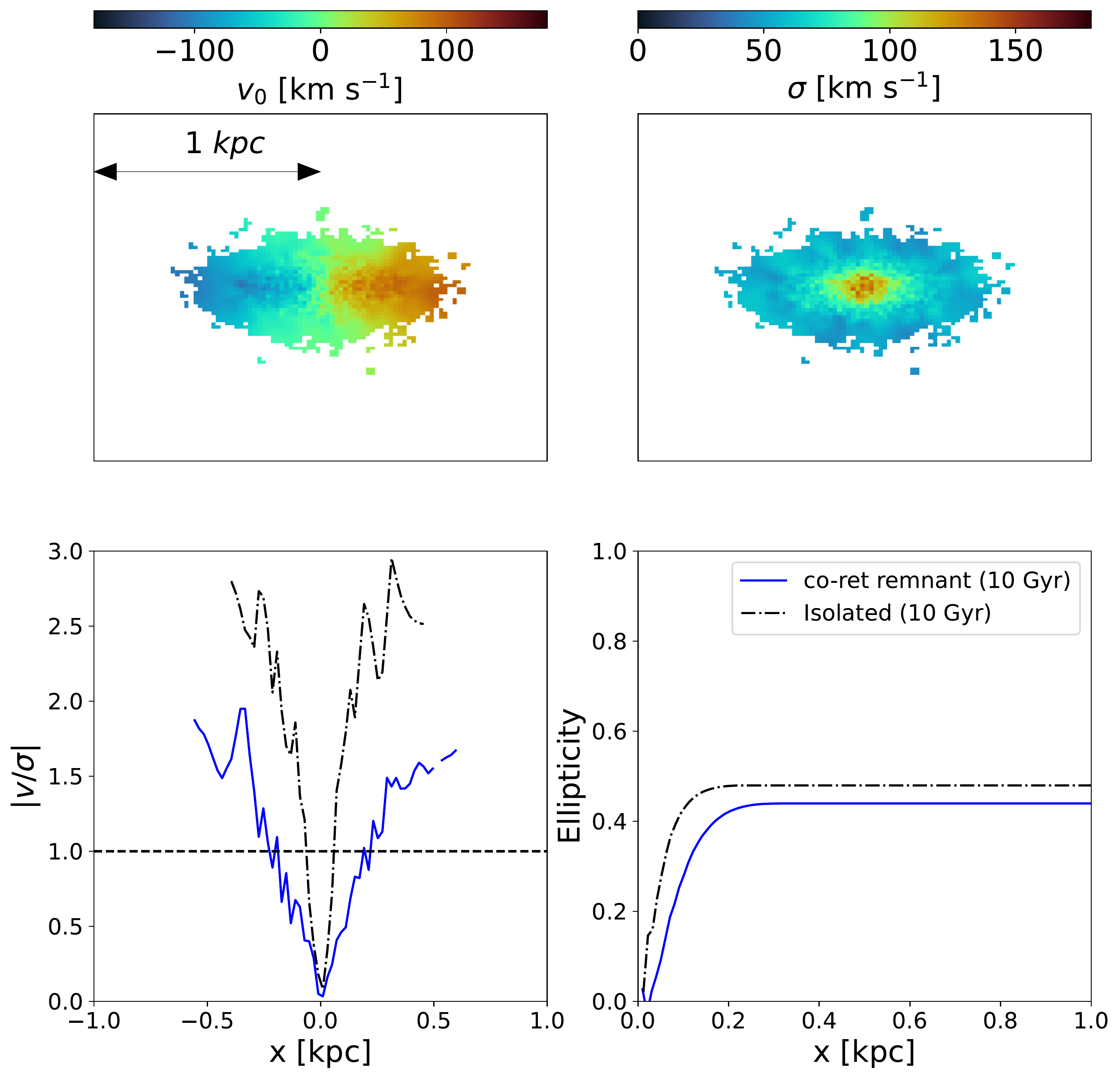}}
    \caption{Same as Figure~\ref{fig:voronoi_co-co}, but for the co-ret merger.}
    \label{fig:voronoi_co-ret}
\end{figure*}
}

\newcommand{\placefigFlat}{
\begin{figure}
    \centering
    \includegraphics[scale=0.4]{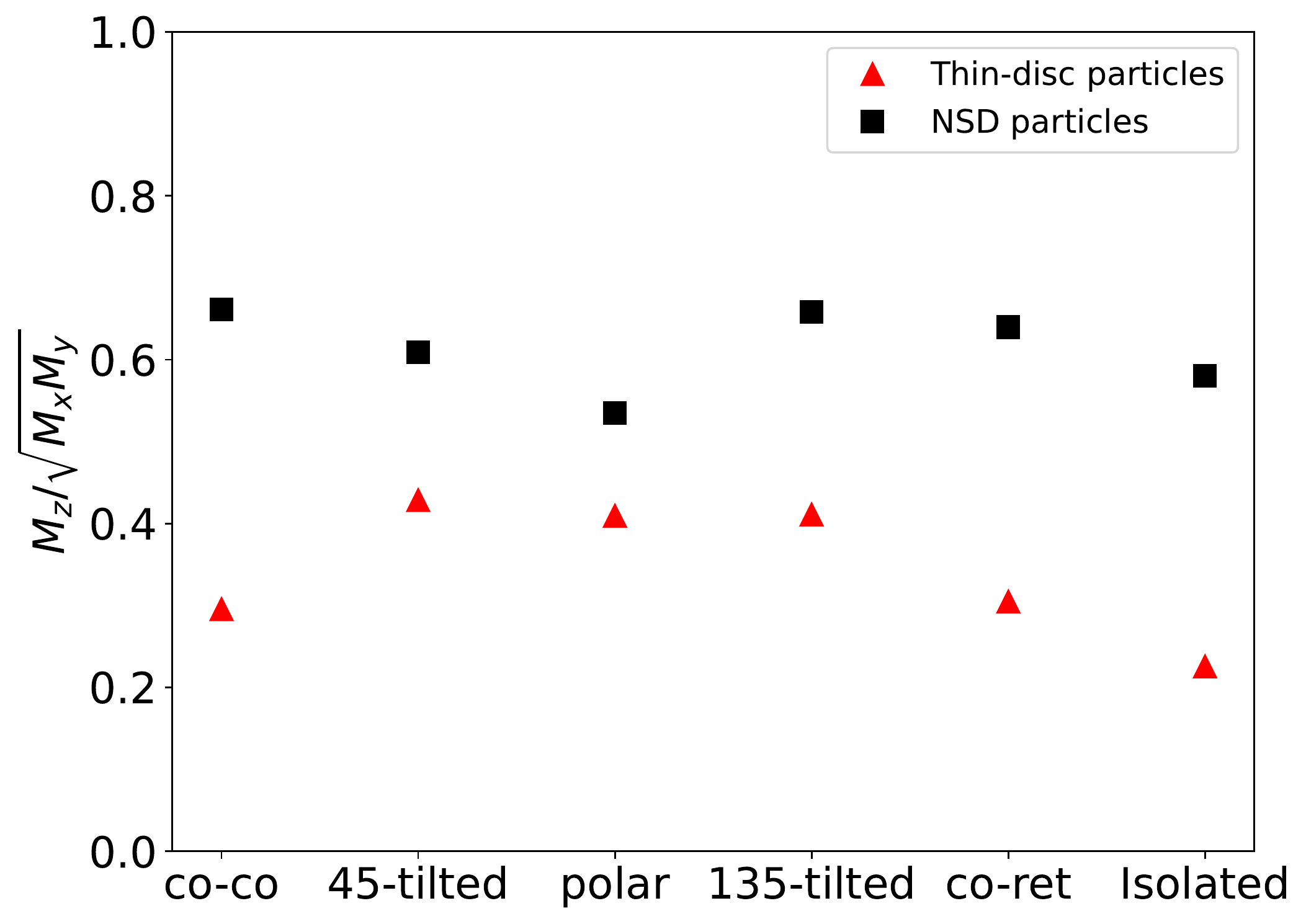}
    \caption{Thickness inside $R_{90}$ for the thin-disc particles (red triangles) and NSD particles (black squares) for our mergers in co-planar co-rotating merger, 45 tilted merger, polar merger, 135 tilted merger, co-planar counter-rotating merger, and for the model in isolation. The thickness for each component has been calculated at 10~Gyr.
    }
    \label{fig:flatness}
\end{figure}
}

\title[The fragility of thin discs in galaxies - II]{The fragility of thin discs in galaxies - II. Thin discs as tracers of the assembly history of galaxies}

\author[P. Galán-de Anta et al.]{Pablo M. Gal\'an-de Anta,$^{1,2}$\thanks{E-mail: pgalandeanta01@qub.ac.uk}
Pedro~R. Capelo,$^{3}$
Eugene Vasiliev,$^{4}$
Massimo Dotti,$^{5,6}$
Marc Sarzi,$^{2}$
\newauthor
Enrico Maria Corsini$^{7,8}$
and Lorenzo Morelli$^{9}$
\\
$^{1}$Astrophysics Research Centre, School of Mathematics and Physics, Queen's University Belfast, Belfast BT7 INN, UK\\
$^{2}$Armagh Observatory and Planetarium, College Hill, Armagh BT61 9DG, UK\\
$^{3}$Center for Theoretical Astrophysics and Cosmology, Institute for Computational Science, University of Zurich,\\ Winterthurerstrasse 190, CH-8057 Zürich, Switzerland\\
$^{4}$Institute of Astronomy, Madingley Road, Cambridge CB3 0HA, UK\\
$^{5}$Dipartimento di Fisica G. Occhialini, Università degli Studi di Milano–Bicocca, Piazza della Scienza 3, I-20126 Milano, Italy\\
$^{6}$INFN, Sezione Milano–Bicocca, Piazza della Scienza 3, I-20126 Milano, Italy\\
$^{7}$Dipartimento di Fisica e Astronomia ``G. Galilei'', Università di Padova, Vicolo dell'Osservatorio 3, I-35122, Padova, Italy\\
$^{8}$INAF-Osservatorio Astronomico di Padova, Vicolo dell'Osservatorio 5, I-35122, Padova, Italy\\
$^{9}$Instituto de Astronomía y Ciencias Planetarias, Universidad de Atacama, Avenida Copayapu 485, Copiapó, Chile\\
}

\date{Accepted XXX. Received YYY; in original form ZZZ}

\pubyear{2023}

\begin{document}
\label{firstpage}
\pagerange{\pageref{firstpage}--\pageref{lastpage}}
\maketitle

\begin{abstract}
Thin galactic discs and nuclear stellar discs (NSDs) are fragile structures that can be easily disturbed by merger events. By studying the age of the stellar populations in present-day discs, we can learn about the assembly history of galaxies and place constraints on their past merger events. Following on the steps of our initial work, we explore the fragility of such disc structures in intermediate-mass-ratio dry encounters using the previously constructed $N$-body model of the Fornax galaxy NGC\,1381 (FCC\,170), which hosts both a thin galactic disc and a NSD. We dismiss major and minor encounters, as the former were previously shown to easily destroy thin-disc structures, whereas the latter take several Hubble times to complete in the specific case of FCC\,170. The kinematics and structure of the thin galactic disc are dramatically altered by the mergers, whereas the NSD shows a remarkable resilience, exhibiting only a smooth increase of its size when compared to the model evolved in isolation. Our results suggest that thin galactic discs are better tracers for intermediate-mass-ratio mergers, while NSDs may be more useful for major encounters. Based on our simulations and previous analysis of the stellar populations, we concluded that FCC\,170 has not experienced any intermediate-mass-ratio dry encounters for at least $\sim$10~Gyr, as indicated by the age of its thin-disc stellar populations.

\end{abstract}

\begin{keywords}
galaxies: elliptical and lenticular, cD -- galaxies: interactions -- galaxies: kinematics and dynamics --  galaxies: structure -- methods: numerical
\end{keywords}



\section{Introduction}

The most widely accepted cosmological paradigm is the $\Lambda$-cold dark matter (CDM) model, wherein galaxies form in a bottom-up fashion \citep[][]{White1978} following the hierarchical growth of the dark matter structure in which they are embedded. In this way smaller galaxies would form first and later grow into larger systems thanks to interactions and merging events, with galaxy mergers also shifting galaxies from disc to spheroidal morphologies \citep[including bulges in disc galaxies;][]{Toomre1977}. Galaxy evolution is therefore dictated by the rate at which galaxies can form stars and merge together, with the latter process being generally more frequent in group and field environments. Here low relative velocities between galaxies favour encounters, as opposed to dense cluster environments, where galaxies evolve mainly through a combination of gravitational and hydrodynamic processes such as tidal interactions and ram-pressure stripping \citep{Moore1996,Angulo2009,Yun2019,Joshi2020,Galan2022}.

While some galaxies may be observed interacting with other galaxies, or exhibiting imprints of past merger events \citep[e.g. tidal features or shells;][]{Malin1980,Malin1983,Schweizer1992,Schweizer1996}, other galaxies do not present any clue of past encounters. However, observational signatures of mergers can dim in time, making necessary in-depth studies of the stellar populations to unveil the assembly history of galaxies \citep[][]{Davison2021,Mazzilli2021}.

\citet{Darg2010} attempted to trace the rate of past merger events by investigating the properties of close galactic pairs and interacting galaxies. These kinds of studies draw on the depth and area covered by recent imaging surveys, finding that merging gas-rich spiral galaxies present intense star formation activity and merging gas-poor ellipticals do not increase their star formation activity. \citet{Eliche2018}, on the other hand, studied the left-over remnant morphologies of major mergers in simulations, finding that S0 galaxies could be the relic of a past major merger. The lack of constraints on the assembly history of galaxies leaves unchecked some of the predictions of the hierarchical paradigm. For example, the formation of the most massive galaxies is not completely understood. Following the $\Lambda$-CDM model, these types of objects should have assembled more recently than observed \citep[][]{Delucia2006,Khochfar2006}.

From a theoretical standpoint, past minor mergers with mass ratios of at least 1:10 (i.e. the ratio of the total mass of the secondary galaxy to that of the primary one is at least 0.1) are thought to have occurred in most galaxies \citep[][]{Ostriker1975,Maller2006,Khochfar2006,Fakhouri2008,Stewart2008,Poole2017,Sotillo2022}. A large fraction of the current galaxy population is expected to have experienced at least one past major encounter of mass ratio $\sim$1:3 in the last $\sim$2--3~Gyr \citep[][]{Bell2006,Lotz2008,Lin2008,Tonnesen2012}, as also confirmed by observations \citep[][]{Lin2004,Barton2007,Woods2007,Yadav2018,Shah2022}. When looking at higher redshifts, violent interactions often affect both the morphology and kinematics of disc galaxies \citep[][]{Hammer2005,Flores2006,Puech2008}. Hence, the fragility of large-scale discs against mergers might be used as a tracer of past merger events \citep{Barnes1992,Hammer2009,Deeley2017}. However, it was shown that discs might also survive mergers of different mass ratios \citep[from 1:3 to <1:10;][]{Abadi2003,Hopkins2009,Purcell2009,Capelo2015}, and merger remnants of wet major mergers could even re-form a galactic disc \citep{Athanassoula2016,CapeloDotti2017,Sparre2017,Peschken2020}. \citet{Toomre1972} laid the first stone by pointing out that mergers can significantly perturb the morphology of discs, even turning them into elliptical galaxies. In fact, the current observed properties of low- and intermediate-mass ellipticals may be the result of past major mergers between spiral galaxies \citep{Hernquist1988,Mihos1994,Barnes1996,Mateo2005,Springel2005,Naab2006}.

In many elliptical galaxies, kpc-scale embedded discs can be easily found, forming a common family of structures in all S0 galaxies \citep{Kormendy1985,Bender1992,Ferrarese2006,Emsellem2007}. Most galaxies also may present disc-like substructures such as nuclear stellar discs (NSDs), first found in Hubble Space Telescope images \citep{Jaffe1994,Bosch1994}. These NSDs reside in the nuclear regions of up to 20 per cent of many kinds of galaxies \citep{Ledo2010} and consist of razor-thin discs of a few hundred parsecs across \citep{Pizzella2002}. The properties of the stellar populations of NSDs were derived for a few galaxies \citep{Corsini2016,Sarzi2016}, with the studies of \citet{Ledo2010} and \citet{Sarzi2015} using, for the first time, NSDs as tracers of the assembly history of galaxies. They performed a simple set of $N$-body mergers consisting of an NSD, a stellar halo, and a central black hole (BH) in interaction with a secondary BH. \citet{Sarzi2015} explored a wide range of mergers, with mass ratios 1:10, 1:5, and 1:1, all in circular orbits and with different inclinations, showing that NSDs may survive some 1:5 encounters and are entirely destroyed in 1:1 mergers.

\citet{Sarzi2015} demonstrated that NSDs may serve as tracers of the most recent merger event, albeit under idealised conditions. In this context, our work aims to investigate the robustness of both kpc-scale thin discs and NSDs against intermediate-mass-ratio mergers, building on our previous research (\citealt{Galan2023}; hereafter \citetalias{Galan2023}). Using FCC\,170 as our primary galaxy and subjecting it to bombardment, we employ a tailored $N$-body model to assess the resilience of its discs to mergers. Taking into account the findings of \citet{Pinna2019}, who indicate that both the thin disc and NSD of FCC\,170 are $\sim$10~Gyr old, we deduce that the age of these stellar components provides a proxy for the look-back time of the last merger event, if the thin discs in our model turn out to be fragile against previous intermediate-mass-ratio encounters.

In Section~\ref{sec:set_run}, we describe how to build the two galaxies used in the merger simulations and provide the details on how to set up the initial conditions of the merger and how to run the simulations. In Section~\ref{sec:survival}, we study the survivability of both the kpc-scale thin disc and NSD against the mergers. Last, in Section~\ref{sec:conclusions} we give our conclusions.

\section{Setting up and running $N$-body simulations}\label{sec:set_run}

\subsection{Initial conditions of isolated galaxies}\label{sec:build_second}

We use the \textsc{Agama} stellar-dynamical toolbox \citep{Vasiliev2019} to create equilibrium models of both the merging galaxies. In \citetalias{Galan2023}, we described in detail the approach for constructing a dynamical model for FCC\,170 that matches the observational constraints. In brief, the galaxy is composed of several components described by distribution functions (DFs) in action space. For any choice of DF parameters, a self-consistent equilibrium model is constructed iteratively by first assuming a gravitational potential, then computing the density generated by the DF of each component in this potential, updating the total potential from the Poisson's equation, and repeating the procedure a few times until convergence. The observable properties of the model (projected density and kinematic maps described by Gauss--Hermite moments) are then compared with the observations, and the parameters of the DFs are varied until a good match is achieved (after many thousand model evaluations). Finally, we create an $N$-body realisation of the best-fitting model and verify that it remains in an equilibrium state for many gigayears, apart from a gradual increase of thickness of the NSD caused by numerical relaxation. We added this model as the primary galaxy.

For the secondary galaxy, we use a simpler approach without extensive parameter search. Namely, we assume that it is a spherical system composed of three components: a central supermassive BH (SMBH), a stellar bulge, and a DM halo. The ratio of SMBH, stellar, and total masses of the secondary galaxy to the primary one is taken to be 1:4, corresponding to the boundary between major and minor mergers (e.g. \citealt{Capelo2015}; see also \citealt{mayer2013} for a discussion). We assume that the DM halo follows an exponentially truncated Navarro--Frenk--White \citep[][]{Navarro1996} profile,

\begin{equation}
    \rho=\rho_0\frac{a}{r}\frac{1}{(1+r/a)^{2}}\exp{\left[-\left(\frac{r}{R_{\rm cutoff}}\right)^{2}\right]},
    \label{eq:NFW}
\end{equation}

\noindent where $a$ is the scale radius, $R_{\rm cutoff}$ is the cutoff radius, and $\rho_{a}$ is the normalisation of the density profile (the total mass is computed by numerical integration). The stellar component follows the \citet{Vaucouleurs1948} profile. We assigned the 3D half-mass radius (or Lagrangian radius at 50 per cent of the mass) of the stellar profile from the scaling relation for spheroidal galaxies \citep[][]{Shen2003}, taking into account that the effective radius of the galaxy is about $\sim$30 per cent smaller than the 3D half-mass radius for a spheroid that follows the \citeauthor{Vaucouleurs1948} profile.  The values of stellar mass and half-mass radius for the secondary (see Table~\ref{tab:secondary_props}) are consistent with the findings for elliptical galaxies reported by \citet{Robertson2006} and also with the fundamental scaling relations found for local galaxies and stellar bulges as studied in \citet{Hon2022}.

\placetabPropertiesNEW

After fixing the structural properties of all galaxy components, we determine the DFs of the stellar bulge and DM halo from the Eddington inversion formula, and use it to assign particle velocities. For consistency with what done in \citetalias{Galan2023}, we choose the individual mass of the stellar and DM particles of the secondary galaxy to be the same as that of the particles of the primary one, i.e. $m_{\star}=7.32\times10^{3}$~M$_{\sun}$ and $m_{\rm DM}=2.48\times10^{5}$~M$_{\sun}$. Hence, the number of particles in the primary galaxy is four times that of the secondary one. The two central SMBHs in our simulations have masses larger than ten times the mass of any individual DM particle. This guarantees that both BHs sink into the centre of the merger remnant by the end of the simulation and are not excessively perturbed by DM particles \citep{Capelo2015}. 
The main values of masses and radii of both the primary FCC\,170 $N$-body model and the secondary spheroid model are tabulated in Table~\ref{tab:secondary_props}.

\subsection{Orbit configuration}\label{sec:orbit}

\placetabMergersNEW

We are interested in reproducing the initial conditions of a merger consistent with the current $\Lambda$-CDM cosmological context. \citet{benson2005} found that the most common type of orbit in cosmological simulations is the parabolic orbit (with eccentricity $e=1$ and binding energy $E=0$), whereas \citet{Khochfar2006} found that about 85 per cent of parabolic encounters have a first pericentric distance larger than 10 per cent of the virial radius\footnote{The virial radius is usually defined as the radius at which the matter density (including both baryonic and DM) is 200 times the critical density of the Universe.} of the primary galaxy.

\placefigMerger

\placefigMergerNSD

The pericentre of a particular orbit is defined by its eccentricity, orbital angular momentum of the primary-secondary system, and virial masses of both the primary and secondary galaxy. The initial separation between the two galaxies is set to $d=R_{\rm cutoff,1}+R_{\rm cutoff,2}$, with $R_{\rm cutoff,1}$ and $R_{\rm cutoff,2}$ the cutoff radii\footnote{The cutoff radius adopts a similar value to the virial radius of the galaxy.} of the DM haloes of the primary and secondary galaxy, respectively. By defining the initial separation larger than the sum of the cutoff radii of the DM haloes, we minimise any tidal effect that could alter the initial morphology of both galaxies during the early stages of the merger. By adjusting the initial orbital angular momentum (i.e. the initial velocities of the galaxies), we can pre-select the first pericentre distance to be about 10--20 per cent of the cutoff radius of the primary. We also varied the initial inclination angle $i$ of the primary with respect to the secondary in order to design different bombardments, ranging from the co-rotating co-planar encounter to the counter-rotating co-planar encounter. We thus set up five encounters with initial inclinations $i=0^{\circ}$, $45^{\circ}$, $90^{\circ}$, $135^{\circ}$, and $180^{\circ}$, with $i=0^{\circ}$ being the co-rotating co-planar encounter and $i=180^{\circ}$ being the counter-rotating co-planar encounter. The name of each merger regarding its inclination angle is tabulated in Table~\ref{tab:merger_props}, along with the main orbital parameters.

\subsection{Details of the runs}

To evolve our $N$-body mergers, we use the code {\textsc{gizmo}} \citep{Hopkins2015}, as explained in section~4.1 of \citetalias{Galan2023}. The softening lengths are equal to those utilised in \citetalias{Galan2023}, i.e. $\varepsilon_{\rm DM}=50$~pc, $\varepsilon_{\rm bulge}=\varepsilon_{\rm thin-disc}=\varepsilon_{\rm thick-disc}=10$~pc, $\varepsilon_{\rm NSD}=5$~pc, and $\varepsilon_{\rm SMBH}=1$~pc, and were chosen to properly resolve the vertical structure of each component in our FCC\,170 $N$-body model. It took about 200 hours of wall-clock time to reach a total integration time of 10~Gyr on 480 processor cores spread between 15 nodes. Same as with the $N$-body FCC\,170 model in isolation, we produce up to 200 snapshots for each merger, equally spaced in time by 50~Myr. The total number of particles for each merger is $N=1.71\times10^{7}$, with $N_{\star}=4.64\times10^{6}$ the number of stellar particles. Prior to the set up of the mergers initial conditions, we initially evolve the secondary galaxy in isolation, proving that it is in equilibrium during the run, with negligible changes in the Lagrangian radii through 10~Gyr.

In Figure~\ref{fig:merger}, we show the evolution of the thin disc of the primary during the co-planar co-rotating merger at different stages: the initial conditions, first, second, and third pericentric passages, and remnant phase at 10~Gyr. We focus on the primary thin disc to show the tidal disruption of the disc throughout the merger. We also show the evolution of the NSD particles in Figure~\ref{fig:merger_NSD}.

Since the secondary galaxy is gradually disrupted by the tidal forces due to the primary, it proves useful to estimate its mass evolution as a function of time. To obtain the bound mass of the secondary galaxy in a given snapshot, we first compute the gravitational potential created by all its particles and then add the kinetic energy of their motion with respect to the galaxy centre, to obtain the total energy $E$ of each particle of the secondary galaxy (still ignoring the effect of the primary). We then remove the particles with $E>0$ and recompute the gravitational potential, repeating this procedure several times until the list of bound particles stabilises. At early times, this list includes almost all the particles of the secondary galaxy. But already after the first pericentric passage, the bound mass considerably drops and eventually decreasing essentially to zero well before the beginning of the ``remnant phase'', which we define to start when the SMBH separation drops below 100~pc. In other words, the SMBH of the secondary galaxy becomes stripped of its surrounding stars and, as a consequence, the efficiency of dynamical friction dramatically decreases: the separation between the two SMBHs remains at a level of a few kiloparsec for several gigayears, before further dropping to essentially the softening length (a few parsec) towards the end of the last (``remnant'') phase, occurring, for all encounters, before 10~Gyr.

In Figure~\ref{fig:distance}, the top panel shows the time evolution of the separation between the central SMBHs of the primary and secondary galaxy for each merger of our suite. The bottom panel shows the total mass loss during the merger. We have not included the mass of the central SMBH of the secondary galaxy in the mass-loss calculation. The first pericentric distances are also tabulated in Table~\ref{tab:merger_props} and are all $\sim0.15 R_{\rm cutoff,1}$. We find that most of the mass loss occurs before the remnant phase, with barely no bound mass at the secondary galaxy during the remnant phase. This suggests that the core of the secondary galaxy gets `naked' before the remnant phase, with just the central secondary SMBH reaching the central region of the galaxy remnant, at scales of the NSD.

\placefigDistances

Additionally, we have also run a set of two mergers for a minor encounter (of mass ratio 1:10) in a co-planar co-rotating orbit and a co-planar counter-rotating orbit (not listed in Tables~\ref{tab:secondary_props}--\ref{tab:merger_props} nor shown in Figure~\ref{fig:distance}). For this set-up, we infer that dry minor mergers with our FCC\,170 $N$-body model do merge in a time-scale that is larger than the current age of the Universe, since dynamical friction is not efficient enough. Consequently, as the secondary galaxy does not reach the centre of the primary galaxy in a reasonable time, these mergers were not utilised as possible tracers of the assembly history in the present work. \citet{Sarzi2015} found that NSDs can withstand any minor merger regardless of the inclination of the orbit, with small to no effects on the final shape of the NSD. In addition, \citet{Sarzi2015} showed that NSDs would not survive major (1:1) mergers, which would no doubt also significantly affect the main stellar disc. Therefore, we focus on exploring the effect exerted by intermediate-mass-ratio mergers on the thin kpc-scale disc and NSD, as it has not been studied in detail yet.

In the next section, we analyse the results for each merger and we check how the encounter affects the thinness and rotation of both the thin kpc-scale disc and NSD. We also compare the results with those of our FCC\,170 $N$-body model evolved in isolation in the same time range.

\placefigFlat

\section{Survivability of thin discs against mergers}\label{sec:survival}

\placefigVoronoiCo

\subsection{Changes in galaxy thickness}\label{sec:shape}


To quantify the morphological changes experienced by the galaxy remnant's thin disc and NSD after the merger, we first align the resulting discs with the galactic plane independently, using the tensor of inertia of the stellar particles within $R_{90}$, the Lagrangian radius enclosing $90$ per cent of the total mass of the galaxy/component, excluding the particles in the outskirts of each disc component. This method was also used by \citet{Joshi2020} and \citet{Pulsoni2020}, and more recently by \citet{Galan2022}. We rotate each galaxy remnant's thin-disc component (thin disc and NSD) independently so that the $z$-axis represents the projected minor axis and the $x$- and $y$-axis are the major axes. Once each component is conveniently aligned with the galactic plane, we quantify the thickness of both the thin disc and the NSD using the mass tensor defined as in \citet{Genel2015}

\begin{equation}
    M_{i}=\frac{\left(\sum_{n}m_{n}x^{2}_{n,i}\right)^{1/2}}{\left(\sum_{n}m_{n}\right)^{1/2}},
    \label{eq:mass_tensor}
\end{equation}

\noindent with the sums performed over all particles inside $R_{90}$, $x_{n,i}$ and $m_{n}$ being the coordinates and mass of each particle, respectively, and $i\in(x,y,z)$. From Equation~\eqref{eq:mass_tensor}, we can infer the thickness of the thin disc and the NSD, given by the axis ratio $M_{z}/\sqrt{M_{x}M_{y}}$, with $M_{i}$ the $i$-th diagonal component of the mass tensor. The thickness accounts for how flat or `discy' a system is: the smaller the thickness, the flatter the system is. Consequently, we can use the thickness parameter to account for how much the flatness of the thin disc and NSD got modified by the mergers.

In Figure~\ref{fig:flatness}, we show the thickness of the remnant for each merger, along with those of the model in isolation (all at 10~Gyr), thin-disc particles (red triangles) and NSD particles (black squares). The thickness is calculated inside $R_{90}$. We observe a considerable increase of the thickness of the thin disc, with respect to the isolated model for the co-co and co-ret mergers (by 31 and 35 per cent, respectively) and a dramatic thickening of the thin disc for the 45-tilted (by 90 per cent), polar (by 82 per cent), and 135-tilted (by 82 per cent) mergers. These numbers suggest that if a past intermediate-mass-ratio merger occurred in this galaxy, it should be imprinted on both the photometry and kinematics of the galaxy. We explore this scenario in detail in Sections~\ref{sec:ellip} and \ref{sec:kinematics}.

The NSD experiences a slight increase of the thickness for four out of five mergers (the polar merger being the exception), with no significant differences amongst them, albeit the co-co, 135-tilted, and co-ret mergers are slightly more disturbing. For the polar merger, the NSD experiences a small contraction in the semi-minor axis too, decreasing its thickness (by $\sim$8 per cent) but slightly increasing its size along the semi-major axis. Most of the particles of the secondary get unbound during the collision, heating the kpc-scale components (thin disc, thick disc, and bulge) and making the primary galaxy to expand. Hence, the secondary galaxy sinks towards the centre of FCC\,170 as a `naked' core with most of its mass in its central SMBH. As the mass of the secondary BH is about 100 times smaller than that of the NSD, the central disc barely notices the presence of the secondary SMBH, just contributing to a modest expansion (or flattening, in the case of the polar merger) of the NSD but being unable to destroy it.

Additionally, for every merger we find that the thin disc and NSD experience a modest misalignment (overall for the tilted mergers) when the two galaxies approach and overall at distances less than $\lesssim10$ kpc at times between $\sim4-6$ Gyr. This misalignment gets compensated at late stages by the thin disc torque, aligning the NSD.

\placefigVoronoiTiltedCo

\subsection{Ellipticity and stellar kinematics of thin discs}\label{sec:ellip}

For each galaxy merger, we also calculate the changes in ellipticity and stellar kinematics of the FCC\,170 $N$-body model at 10~Gyr to further compare how much the rotation has diminished and how much the shape has been modified, in particular for both the thin disc and NSD. After aligning the galaxy remnant with the galactic plane through the inertia tensor \citep{Joshi2020,Pulsoni2020,Galan2022}, we further design a 2D grid of the projected edge-on galactic thin-disc and NSD remnants by binning the thin disc and NSD inside a 20$\times$20~kpc box and 2$\times$2~kpc box, respectively. We choose bin sizes of 200~pc and 20~pc for the thin disc and NSD, respectively. We then design an adaptive-bandwidth histogram using a kernel density estimate from $K$ nearest neighbours using $K=100$ in each mock image to recover the velocity dispersion of the thin disc and NSD of our galaxy remnants even at pixels with few points. Once we have the 2D mock images of both the thin disc and NSD remnants, we compute the ellipticity at different radii using equation~(1) from \citet{Emsellem2007}

\begin{equation}
    \varepsilon=1-\sqrt{\frac{\langle z^{2}\rangle}{\langle x^{2}\rangle}},
    \label{eq:ellip}
\end{equation}

\noindent where $\langle z^{2} \rangle$ is the quadratic mass-weighted coordinate along the semi-minor axis, defined as ${\langle z^{2} \rangle=\sum_{n}m_{n}z^{2}_{n}/\sum_{n}m_{n}}$ sum over $n$ bins, and $\langle x^{2} \rangle$ the quadratic mass-weighted coordinate along the semi-major axis. For an infinitely flat disc and a perfect sphere, it is $\varepsilon\sim1$ and $\varepsilon=0$, respectively.

The top panels of Figures~\ref{fig:voronoi_co-co}, \ref{fig:voronoi_tilted-co}, \ref{fig:voronoi_polar}, \ref{fig:voronoi_tiled-ret}, and \ref{fig:voronoi_co-ret} show the mean velocity $v_{0}$ and velocity dispersion $\sigma$ for the thin-disc particles (two top-left panels) and NSD particles (two top-right panels) of the merger remnant for the co-co, 45-tilted, polar, 135-tilted, and co-ret merger, respectively. In the bottom panels, we show the profiles of $|v/\sigma|$ and ellipticity along the major axis with blue solid lines. We also show the values of $|v/\sigma|$ and ellipticity for the FCC\,170 $N$-body model in isolation at the same time with black dotted-dashed lines.

The shape of the thin disc gets significantly distorted by the 45-tilted, polar, and 135-tilted encounters, diminishing its ellipticity (turning into a thicker disc) on a plateau at radii larger than $\sim$3~kpc with respect to the model in isolation at 10~Gyr. The relative decrements in ellipticity at the plateau with respect to the model in isolation is 55, 59, and 43 per cent, for the 45-tilted, polar, and 135-tilted merger, respectively. On the other hand, the co-co and co-ret mergers decrease the ellipticity of the thin disc by about 7 per cent. Regarding the NSD, the changes in ellipticity are quite similar for the co-co and co-ret mergers, with the ellipticities decreasing by 12 and 8 per cent, respectively. For the 45-tilted, polar, and 135-tilted mergers, the ellipticity diminished by 9, 1, and 15 per cent, respectively.

\placefigVoronoiPolar

Regarding the kinematics, each merger substantially decreases the rotation of the thin disc with the tilted and polar mergers being the most effective. For the 45-tilted, polar, and 135-tilted encounters, while the disc gets twisted and becomes thicker, its rotation significantly decreases with respect to the model in isolation, overall at radii greater than $\sim$3~kpc, where the structure of the thin disc gets thicker. Regarding the kinematics of the NSD, the mergers that affect the rotation the least are the co-co and polar mergers, with a modest decrease of the NSD rotation. The co-ret, 45-tilted, and 135-tilted mergers affect the most the NSD rotation. 

As the secondary SMBH reaches the centre of FCC\,170 as a `naked' SMBH, with a mass that is negligible compared to that of the NSD, the mergers do not destroy the NSD, albeit they decrease its rotation while slightly decreasing its ellipticity. These findings are different from those found by \citet{Sarzi2015}. In their case, for a 1:5 merger at different orbit inclinations, only for the co-planar co-rotating orbit the NSD got totally destroyed. On the contrary, in our case, regardless of the orbit, the NSD only experiences an expansion without being entirely destroyed. In their experiments, however, the mass of the secondary SMBH was comparable to that of the NSD, whereas in our case the mass of the secondary SMBH is significantly ($\sim100$ times) smaller than the mass of the NSD. Taken together with the outcome of our simulations, this suggests that NSDs in galaxies bombarded by an intermediate-mass-ratio-merger perturber might be less resilient if the mass of the secondary SMBH is of the order of that of the NSD.

Regarding numerical heating on the NSD, as also discussed in \citetalias{Galan2023}, the results show that the changes in the ellipticity of the NSD are larger when the disc is exposed to a galaxy merger than in the isolated case, except for the polar merger, in which the decrement in ellipticity is comparable to that shown for the isolated case. In general, the NSDs of the merger remnants exhibit a smaller ellipticity than in the isolated scenario, and the polar remnant shows comparable values, suggesting that the effect of numerical heating is quantitatively similar to the effects of the merger remnants.

Although in our simulations we have not included gas particles, we can expect that the outcome of a gas-rich merger is likely to include central star formation and addition of stellar disc populations \citep[e.g.][]{Abadi2003,Hopkins2009,VanWassenhove2014,Capelo2015,CapeloDotti2017}. The precise reconstruction of the star-formation history of NSDs \citep[e.g.][]{Pinna2019} is a promising avenue to constrain the occurrence of such last wet merger events.

\subsection{Kinematics comparison: merger remnants versus the isolated model}\label{sec:kinematics}

Last, we check the similarity between the kinematics of the entire merger remnants and FCC\,170 $N$-body model after 10~Gyr of evolution in isolation. Such a comparison allows to quantify how much the merger events we considered would affect the overall stellar morphology and kinematics of an initially unperturbed FCC~170-like system, and to whether or not such differences with respect to the isolated model for would be significant from an observer perspective.

To produce kinematic 2D maps for both the model in isolation and each merger remnant, we follow the procedure outlined in section~3.2 of \citetalias{Galan2023} to finally obtain the mean velocity $v_{0}$ and velocity dispersion $\sigma$. In Figure~\ref{fig:kinematics_mergers}, we show the surface stellar luminosity, which, as in \citetalias{Galan2023}, is reproduced as an integral of the DF over the velocity. We also plot the mean velocity and the velocity dispersion for the model in isolation and of our merger remnants. Whereas in the mock images of Figures~\ref{fig:voronoi_co-co}--\ref{fig:voronoi_co-ret} we show just a single component independently (thin disc or NSD), in Figure~\ref{fig:kinematics_mergers} we display all the stellar particles. Figure~\ref{fig:kinematics_mergers} shows that, although the surface stellar luminosity distribution and velocity maps are not dramatically different after an intermediate-mass-ratio merger, the stellar velocity dispersion map is radically different from that of a galaxy with a prominent thin-disc structure. Such an encounter would indeed lead to shallower velocity dispersion gradients and larger average values of the velocity dispersion. In all considered merging scenarios, the tell-tale signature of a thin disc (i.e. the low-velocity dispersion values near the equatorial plane outside the bulge-dominated regions) would be erased. On the other hand, all merger remnants show a central dip in velocity dispersion (not easily visible in the figures because of the edge-on projections), thus preserving the kinematic signature of the NSD, in agreement with the analysis made in Section~\ref{sec:ellip}.

\placefigVoronoiTiltedRet

If FCC\,170 had experienced an intermediate-mass-ratio dry merger, our results suggest that the kinematic maps of the corresponding galaxy remnants would not match those of our isolated numerical model for FCC\,170, which in turn was shown in \citetalias{Galan2023} to be kinematically consistent with the actual MUSE observations for this galaxy. In turn, since the measurements of \citet{Pinna2019} indicate that both the NSD and thin disc of FCC~170 have passively evolving stellar populations that are at least 10~Gyr old, we infer that FCC\,170 did not experience any major or intermediate-mass-ratio merger events over this period of time. Deep imaging of the Fornax cluster by \citet{Iodice2019} further show no evidence of tidal tails or shells in the outskirts of FCC\,170, suggesting  that also relatively minor mergers did not take place in the recent past of this object.

The absence of any kinematic or photometric signature of significant gravitational interactions for FCC\,170 is in agreement with the notion that galaxies in clusters have a low probability of merging with other galaxies due to their relatively high velocities \citep{Serra2017}, as suggested by the survival of discs in simulated cluster galaxies \citep{Joshi2020, Galan2022}.

\section{Conclusions}\label{sec:conclusions}

Making use of the $N$-body model of FCC\,170 created in \citetalias{Galan2023}, we run a set of 1:4 dry mergers in parabolic orbits with different inclinations to study the resilience of the kpc-scale thin disc and NSD against such encounters.

We find that the thin, kpc-scale disc gets destroyed in the polar and tilted encounters, whereas it experiences a modest expansion in the co-planar, co-rotating, and counter-rotating orbits, indicating that thin galactic discs may be resilient against co-planar, intermediate-mass-ratio dry encounters. On the other hand, the NSD of FCC\,170 appears to be quite resilient, surviving all the encounters in our suite of simulations while increasing its thickness and still being observable in the kinematics of the galaxy. This suggests that NSDs in galaxies could be more resilient than previously thought against intermediate-mass-ratio encounters, in particular if they are rather more massive than the central SMBH of the secondary galaxy \citep[and not simply assumed to be of similar mass, as done in][]{Sarzi2015}, which in turn makes them less susceptible to the interaction.

Our results also strongly indicate that, according to the latest estimates of the stellar ages of FCC\,170, it is rather unlikely that this galaxy has experienced a past intermediate-mass-ratio event in the last 10~Gyr, as inferred from the kinematics of the merger remnant, and in good agreement with the cluster environment where the galaxy is embedded, in which direct collisions between galaxies are a rare event.

\section*{Acknowledgements}

\placefigVoronoiRet

\placefigKinem

This work was performed on the OzSTAR national facility at Swinburne University of Technology. The OzSTAR programme receives funding in part from the Astronomy National Collaborative Research Infrastructure Strategy (NCRIS) allocation provided by the Australian Government. We are grateful for use of the computing resources from the Northern Ireland High Performance Computing (NI-HPC) service funded by the Engineering and Physical Sciences Research Council (EPSRC) (EP/T022175). Enrico Maria Corsini acknowledges support by Padua University grants DOR 2019-2022 and by Italian Ministry for Education University and Research (MIUR) grant PRIN 2017 20173ML3WW-001.

\section*{Data Availability Statement}

The data underlying this article can be made available upon request. The models results can be reproduced using publicly available codes.



\scalefont{0.94}
\setlength{\bibhang}{1.6em}
\setlength\labelwidth{0.0em}
\bibliographystyle{mnras}
\bibliography{paper} 

\begin{thebibliography}{}
\makeatletter
\relax
\def\mn@urlcharsother{\let\do\@makeother \do\$\do\&\do\#\do\^\do\_\do\%\do\~}
\def\mn@doi{\begingroup\mn@urlcharsother \@ifnextchar [ {\mn@doi@}
  {\mn@doi@[]}}
\def\mn@doi@[#1]#2{\def\@tempa{#1}\ifx\@tempa\@empty \href
  {http://dx.doi.org/#2} {doi:#2}\else \href {http://dx.doi.org/#2} {#1}\fi
  \endgroup}
\def\mn@eprint#1#2{\mn@eprint@#1:#2::\@nil}
\def\mn@eprint@arXiv#1{\href {http://arxiv.org/abs/#1} {{\tt arXiv:#1}}}
\def\mn@eprint@dblp#1{\href {http://dblp.uni-trier.de/rec/bibtex/#1.xml}
  {dblp:#1}}
\def\mn@eprint@#1:#2:#3:#4\@nil{\def\@tempa {#1}\def\@tempb {#2}\def\@tempc
  {#3}\ifx \@tempc \@empty \let \@tempc \@tempb \let \@tempb \@tempa \fi \ifx
  \@tempb \@empty \def\@tempb {arXiv}\fi \@ifundefined
  {mn@eprint@\@tempb}{\@tempb:\@tempc}{\expandafter \expandafter \csname
  mn@eprint@\@tempb\endcsname \expandafter{\@tempc}}}

\bibitem[\protect\citeauthoryear{{Abadi}, {Navarro}, {Steinmetz}  \&
  {Eke}}{{Abadi} et~al.}{2003}]{Abadi2003}
{Abadi} M.~G.,  {Navarro} J.~F.,  {Steinmetz} M.,   {Eke} V.~R.,  2003, \mn@doi
  [\apj] {10.1086/378316}, \href
  {https://ui.adsabs.harvard.edu/abs/2003ApJ...597...21A} {597, 21}

\bibitem[\protect\citeauthoryear{{Angulo}, {Lacey}, {Baugh}  \&
  {Frenk}}{{Angulo} et~al.}{2009}]{Angulo2009}
{Angulo} R.~E.,  {Lacey} C.~G.,  {Baugh} C.~M.,   {Frenk} C.~S.,  2009, \mn@doi
  [\mnras] {10.1111/j.1365-2966.2009.15333.x}, \href
  {https://ui.adsabs.harvard.edu/abs/2009MNRAS.399..983A} {399, 983}

\bibitem[\protect\citeauthoryear{{Athanassoula}, {Rodionov}, {Peschken}  \&
  {Lambert}}{{Athanassoula} et~al.}{2016}]{Athanassoula2016}
{Athanassoula} E.,  {Rodionov} S.~A.,  {Peschken} N.,   {Lambert} J.~C.,  2016,
  \mn@doi [\apj] {10.3847/0004-637X/821/2/90}, \href
  {https://ui.adsabs.harvard.edu/abs/2016ApJ...821...90A} {821, 90}

\bibitem[\protect\citeauthoryear{{Barnes} \& {Hernquist}}{{Barnes} \&
  {Hernquist}}{1992}]{Barnes1992}
{Barnes} J.~E.,  {Hernquist} L.,  1992, \mn@doi [\araa]
  {10.1146/annurev.aa.30.090192.003421}, \href
  {https://ui.adsabs.harvard.edu/abs/1992ARA&A..30..705B} {30, 705}

\bibitem[\protect\citeauthoryear{{Barnes} \& {Hernquist}}{{Barnes} \&
  {Hernquist}}{1996}]{Barnes1996}
{Barnes} J.~E.,  {Hernquist} L.,  1996, \mn@doi [\apj] {10.1086/177957}, \href
  {https://ui.adsabs.harvard.edu/abs/1996ApJ...471..115B} {471, 115}

\bibitem[\protect\citeauthoryear{{Barton}, {Arnold}, {Zentner}, {Bullock}  \&
  {Wechsler}}{{Barton} et~al.}{2007}]{Barton2007}
{Barton} E.~J.,  {Arnold} J.~A.,  {Zentner} A.~R.,  {Bullock} J.~S.,
  {Wechsler} R.~H.,  2007, \mn@doi [\apj] {10.1086/522620}, \href
  {https://ui.adsabs.harvard.edu/abs/2007ApJ...671.1538B} {671, 1538}

\bibitem[\protect\citeauthoryear{{Bell}, {Phleps}, {Somerville}, {Wolf},
  {Borch}  \& {Meisenheimer}}{{Bell} et~al.}{2006}]{Bell2006}
{Bell} E.~F.,  {Phleps} S.,  {Somerville} R.~S.,  {Wolf} C.,  {Borch} A.,
  {Meisenheimer} K.,  2006, \mn@doi [\apj] {10.1086/508408}, \href
  {https://ui.adsabs.harvard.edu/abs/2006ApJ...652..270B} {652, 270}

\bibitem[\protect\citeauthoryear{{Bender}, {Burstein}  \& {Faber}}{{Bender}
  et~al.}{1992}]{Bender1992}
{Bender} R.,  {Burstein} D.,   {Faber} S.~M.,  1992, \mn@doi [\apj]
  {10.1086/171940}, \href
  {https://ui.adsabs.harvard.edu/abs/1992ApJ...399..462B} {399, 462}

\bibitem[\protect\citeauthoryear{{Benson}}{{Benson}}{2005}]{benson2005}
{Benson} A.~J.,  2005, \mn@doi [\mnras] {10.1111/j.1365-2966.2005.08788.x},
  \href {https://ui.adsabs.harvard.edu/abs/2005MNRAS.358..551B} {358, 551}

\bibitem[\protect\citeauthoryear{{Capelo} \& {Dotti}}{{Capelo} \&
  {Dotti}}{2017}]{CapeloDotti2017}
{Capelo} P.~R.,  {Dotti} M.,  2017, \mn@doi [\mnras] {10.1093/mnras/stw2872},
  \href {https://ui.adsabs.harvard.edu/abs/2017MNRAS.465.2643C} {465, 2643}

\bibitem[\protect\citeauthoryear{{Capelo}, {Volonteri}, {Dotti}, {Bellovary},
  {Mayer}  \& {Governato}}{{Capelo} et~al.}{2015}]{Capelo2015}
{Capelo} P.~R.,  {Volonteri} M.,  {Dotti} M.,  {Bellovary} J.~M.,  {Mayer} L.,
   {Governato} F.,  2015, \mn@doi [\mnras] {10.1093/mnras/stu2500}, \href
  {https://ui.adsabs.harvard.edu/abs/2015MNRAS.447.2123C} {447, 2123}

\bibitem[\protect\citeauthoryear{{Corsini}, {Morelli}, {Pastorello}, {Dalla
  Bont{\`a}}, {Pizzella}  \& {Portaluri}}{{Corsini} et~al.}{2016}]{Corsini2016}
{Corsini} E.~M.,  {Morelli} L.,  {Pastorello} N.,  {Dalla Bont{\`a}} E.,
  {Pizzella} A.,   {Portaluri} E.,  2016, \mn@doi [\mnras]
  {10.1093/mnras/stv2864}, \href
  {https://ui.adsabs.harvard.edu/abs/2016MNRAS.457.1198C} {457, 1198}

\bibitem[\protect\citeauthoryear{{Darg} et~al.,}{{Darg}
  et~al.}{2010}]{Darg2010}
{Darg} D.~W.,  et~al., 2010, \mn@doi [\mnras]
  {10.1111/j.1365-2966.2009.15686.x}, \href
  {https://ui.adsabs.harvard.edu/abs/2010MNRAS.401.1043D} {401, 1043}

\bibitem[\protect\citeauthoryear{{Davison} et~al.,}{{Davison}
  et~al.}{2021}]{Davison2021}
{Davison} T.~A.,  et~al., 2021, \mn@doi [\mnras] {10.1093/mnras/stab162}, \href
  {https://ui.adsabs.harvard.edu/abs/2021MNRAS.502.2296D} {502, 2296}

\bibitem[\protect\citeauthoryear{{De Lucia}, {Springel}, {White}, {Croton}  \&
  {Kauffmann}}{{De Lucia} et~al.}{2006}]{Delucia2006}
{De Lucia} G.,  {Springel} V.,  {White} S. D.~M.,  {Croton} D.,   {Kauffmann}
  G.,  2006, \mn@doi [\mnras] {10.1111/j.1365-2966.2005.09879.x}, \href
  {https://ui.adsabs.harvard.edu/abs/2006MNRAS.366..499D} {366, 499}

\bibitem[\protect\citeauthoryear{{Deeley} et~al.,}{{Deeley}
  et~al.}{2017}]{Deeley2017}
{Deeley} S.,  et~al., 2017, \mn@doi [\mnras] {10.1093/mnras/stx441}, \href
  {https://ui.adsabs.harvard.edu/abs/2017MNRAS.467.3934D} {467, 3934}

\bibitem[\protect\citeauthoryear{{Di Matteo}, {Springel}  \& {Hernquist}}{{Di
  Matteo} et~al.}{2005}]{Mateo2005}
{Di Matteo} T.,  {Springel} V.,   {Hernquist} L.,  2005, \mn@doi [\nat]
  {10.1038/nature03335}, \href
  {https://ui.adsabs.harvard.edu/abs/2005Natur.433..604D} {433, 604}

\bibitem[\protect\citeauthoryear{{Eliche-Moral}, {Rodr{\'\i}guez-P{\'e}rez},
  {Borlaff}, {Querejeta}  \& {Tapia}}{{Eliche-Moral} et~al.}{2018}]{Eliche2018}
{Eliche-Moral} M.~C.,  {Rodr{\'\i}guez-P{\'e}rez} C.,  {Borlaff} A.,
  {Querejeta} M.,   {Tapia} T.,  2018, \mn@doi [\aap]
  {10.1051/0004-6361/201832911}, \href
  {https://ui.adsabs.harvard.edu/abs/2018A&A...617A.113E} {617, A113}

\bibitem[\protect\citeauthoryear{{Emsellem} et~al.,}{{Emsellem}
  et~al.}{2007}]{Emsellem2007}
{Emsellem} E.,  et~al., 2007, \mn@doi [\mnras]
  {10.1111/j.1365-2966.2007.11752.x}, \href
  {https://ui.adsabs.harvard.edu/abs/2007MNRAS.379..401E} {379, 401}

\bibitem[\protect\citeauthoryear{{Fakhouri} \& {Ma}}{{Fakhouri} \&
  {Ma}}{2008}]{Fakhouri2008}
{Fakhouri} O.,  {Ma} C.-P.,  2008, \mn@doi [\mnras]
  {10.1111/j.1365-2966.2008.13075.x}, \href
  {https://ui.adsabs.harvard.edu/abs/2008MNRAS.386..577F} {386, 577}

\bibitem[\protect\citeauthoryear{{Ferrarese} et~al.,}{{Ferrarese}
  et~al.}{2006}]{Ferrarese2006}
{Ferrarese} L.,  et~al., 2006, \mn@doi [\apjs] {10.1086/501350}, \href
  {https://ui.adsabs.harvard.edu/abs/2006ApJS..164..334F} {164, 334}

\bibitem[\protect\citeauthoryear{{Flores}, {Hammer}, {Puech}, {Amram}  \&
  {Balkowski}}{{Flores} et~al.}{2006}]{Flores2006}
{Flores} H.,  {Hammer} F.,  {Puech} M.,  {Amram} P.,   {Balkowski} C.,  2006,
  \mn@doi [\aap] {10.1051/0004-6361:20054217}, \href
  {https://ui.adsabs.harvard.edu/abs/2006A&A...455..107F} {455, 107}

\bibitem[\protect\citeauthoryear{{Gal{\'a}n-de Anta} et~al.,}{{Gal{\'a}n-de
  Anta} et~al.}{2022}]{Galan2022}
{Gal{\'a}n-de Anta} P.~M.,  et~al., 2022, \mn@doi [\mnras]
  {10.1093/mnras/stac3061}, \href
  {https://ui.adsabs.harvard.edu/abs/2022MNRAS.517.5992G} {517, 5992}

\bibitem[\protect\citeauthoryear{{Gal{\'a}n-de Anta} et~al.,}{{Gal{\'a}n-de
  Anta} et~al.}{2023}]{Galan2023}
{Gal{\'a}n-de Anta} P.~M.,  et~al., 2023, \mn@doi [\mnras]
  {10.1093/mnras/stad419}, \href
  {https://ui.adsabs.harvard.edu/abs/2023MNRAS.520.4490G} {520, 4490}

\bibitem[\protect\citeauthoryear{{Genel}, {Fall}, {Hernquist}, {Vogelsberger},
  {Snyder}, {Rodriguez-Gomez}, {Sijacki}  \& {Springel}}{{Genel}
  et~al.}{2015}]{Genel2015}
{Genel} S.,  {Fall} S.~M.,  {Hernquist} L.,  {Vogelsberger} M.,  {Snyder}
  G.~F.,  {Rodriguez-Gomez} V.,  {Sijacki} D.,   {Springel} V.,  2015, \mn@doi
  [\apjl] {10.1088/2041-8205/804/2/L40}, \href
  {https://ui.adsabs.harvard.edu/abs/2015ApJ...804L..40G} {804, L40}

\bibitem[\protect\citeauthoryear{{Hammer}, {Flores}, {Elbaz}, {Zheng}, {Liang}
  \& {Cesarsky}}{{Hammer} et~al.}{2005}]{Hammer2005}
{Hammer} F.,  {Flores} H.,  {Elbaz} D.,  {Zheng} X.~Z.,  {Liang} Y.~C.,
  {Cesarsky} C.,  2005, \mn@doi [\aap] {10.1051/0004-6361:20041471}, \href
  {https://ui.adsabs.harvard.edu/abs/2005A&A...430..115H} {430, 115}

\bibitem[\protect\citeauthoryear{{Hammer}, {Flores}, {Yang}, {Athanassoula},
  {Puech}, {Rodrigues}  \& {Peirani}}{{Hammer} et~al.}{2009}]{Hammer2009}
{Hammer} F.,  {Flores} H.,  {Yang} Y.~B.,  {Athanassoula} E.,  {Puech} M.,
  {Rodrigues} M.,   {Peirani} S.,  2009, \mn@doi [\aap]
  {10.1051/0004-6361:200810488}, \href
  {https://ui.adsabs.harvard.edu/abs/2009A&A...496..381H} {496, 381}

\bibitem[\protect\citeauthoryear{{Hernquist} \& {Quinn}}{{Hernquist} \&
  {Quinn}}{1988}]{Hernquist1988}
{Hernquist} L.,  {Quinn} P.~J.,  1988, \mn@doi [\apj] {10.1086/166592}, \href
  {https://ui.adsabs.harvard.edu/abs/1988ApJ...331..682H} {331, 682}

\bibitem[\protect\citeauthoryear{{Hon}, {Graham}  \& {Sahu}}{{Hon}
  et~al.}{2023}]{Hon2022}
{Hon} D. S.~H.,  {Graham} A.~W.,   {Sahu} N.,  2023, \mn@doi [\mnras]
  {10.1093/mnras/stac3704}, \href
  {https://ui.adsabs.harvard.edu/abs/2023MNRAS.519.4651H} {519, 4651}

\bibitem[\protect\citeauthoryear{{Hopkins}}{{Hopkins}}{2015}]{Hopkins2015}
{Hopkins} P.~F.,  2015, \mn@doi [\mnras] {10.1093/mnras/stv195}, \href
  {https://ui.adsabs.harvard.edu/abs/2015MNRAS.450...53H} {450, 53}

\bibitem[\protect\citeauthoryear{{Hopkins}, {Cox}, {Younger}  \&
  {Hernquist}}{{Hopkins} et~al.}{2009}]{Hopkins2009}
{Hopkins} P.~F.,  {Cox} T.~J.,  {Younger} J.~D.,   {Hernquist} L.,  2009,
  \mn@doi [\apj] {10.1088/0004-637X/691/2/1168}, \href
  {https://ui.adsabs.harvard.edu/abs/2009ApJ...691.1168H} {691, 1168}

\bibitem[\protect\citeauthoryear{{Iodice} et~al.,}{{Iodice}
  et~al.}{2019}]{Iodice2019}
{Iodice} E.,  et~al., 2019, \mn@doi [\aap] {10.1051/0004-6361/201833741}, \href
  {https://ui.adsabs.harvard.edu/abs/2019A&A...623A...1I} {623, A1}

\bibitem[\protect\citeauthoryear{{Jaffe}, {Ford}, {O'Connell}, {van den Bosch}
  \& {Ferrarese}}{{Jaffe} et~al.}{1994}]{Jaffe1994}
{Jaffe} W.,  {Ford} H.~C.,  {O'Connell} R.~W.,  {van den Bosch} F.~C.,
  {Ferrarese} L.,  1994, \mn@doi [\aj] {10.1086/117178}, \href
  {https://ui.adsabs.harvard.edu/abs/1994AJ....108.1567J} {108, 1567}

\bibitem[\protect\citeauthoryear{{Joshi}, {Pillepich}, {Nelson}, {Marinacci},
  {Springel}, {Rodriguez-Gomez}, {Vogelsberger}  \& {Hernquist}}{{Joshi}
  et~al.}{2020}]{Joshi2020}
{Joshi} G.~D.,  {Pillepich} A.,  {Nelson} D.,  {Marinacci} F.,  {Springel} V.,
  {Rodriguez-Gomez} V.,  {Vogelsberger} M.,   {Hernquist} L.,  2020, \mn@doi
  [\mnras] {10.1093/mnras/staa1668}, \href
  {https://ui.adsabs.harvard.edu/abs/2020MNRAS.496.2673J} {496, 2673}

\bibitem[\protect\citeauthoryear{{Khochfar} \& {Burkert}}{{Khochfar} \&
  {Burkert}}{2006}]{Khochfar2006}
{Khochfar} S.,  {Burkert} A.,  2006, \mn@doi [\aap]
  {10.1051/0004-6361:20053241}, \href
  {https://ui.adsabs.harvard.edu/abs/2006A&A...445..403K} {445, 403}

\bibitem[\protect\citeauthoryear{{Kormendy}}{{Kormendy}}{1985}]{Kormendy1985}
{Kormendy} J.,  1985, \mn@doi [\apj] {10.1086/163350}, \href
  {https://ui.adsabs.harvard.edu/abs/1985ApJ...295...73K} {295, 73}

\bibitem[\protect\citeauthoryear{{Ledo}, {Sarzi}, {Dotti}, {Khochfar}  \&
  {Morelli}}{{Ledo} et~al.}{2010}]{Ledo2010}
{Ledo} H.~R.,  {Sarzi} M.,  {Dotti} M.,  {Khochfar} S.,   {Morelli} L.,  2010,
  \mn@doi [\mnras] {10.1111/j.1365-2966.2010.16990.x}, \href
  {https://ui.adsabs.harvard.edu/abs/2010MNRAS.407..969L} {407, 969}

\bibitem[\protect\citeauthoryear{{Lin} et~al.,}{{Lin} et~al.}{2004}]{Lin2004}
{Lin} L.,  et~al., 2004, \mn@doi [\apjl] {10.1086/427183}, \href
  {https://ui.adsabs.harvard.edu/abs/2004ApJ...617L...9L} {617, L9}

\bibitem[\protect\citeauthoryear{{Lin} et~al.,}{{Lin} et~al.}{2008}]{Lin2008}
{Lin} L.,  et~al., 2008, \mn@doi [\apj] {10.1086/587928}, \href
  {https://ui.adsabs.harvard.edu/abs/2008ApJ...681..232L} {681, 232}

\bibitem[\protect\citeauthoryear{{Lotz} et~al.,}{{Lotz}
  et~al.}{2008}]{Lotz2008}
{Lotz} J.~M.,  et~al., 2008, \mn@doi [\apj] {10.1086/523659}, \href
  {https://ui.adsabs.harvard.edu/abs/2008ApJ...672..177L} {672, 177}

\bibitem[\protect\citeauthoryear{{Malin} \& {Carter}}{{Malin} \&
  {Carter}}{1980}]{Malin1980}
{Malin} D.~F.,  {Carter} D.,  1980, \mn@doi [\nat] {10.1038/285643a0}, \href
  {https://ui.adsabs.harvard.edu/abs/1980Natur.285..643M} {285, 643}

\bibitem[\protect\citeauthoryear{{Malin} \& {Carter}}{{Malin} \&
  {Carter}}{1983}]{Malin1983}
{Malin} D.~F.,  {Carter} D.,  1983, \mn@doi [\apj] {10.1086/161467}, \href
  {https://ui.adsabs.harvard.edu/abs/1983ApJ...274..534M} {274, 534}

\bibitem[\protect\citeauthoryear{{Maller}, {Katz}, {Kere{\v{s}}}, {Dav{\'e}}
  \& {Weinberg}}{{Maller} et~al.}{2006}]{Maller2006}
{Maller} A.~H.,  {Katz} N.,  {Kere{\v{s}}} D.,  {Dav{\'e}} R.,   {Weinberg}
  D.~H.,  2006, \mn@doi [\apj] {10.1086/503319}, \href
  {https://ui.adsabs.harvard.edu/abs/2006ApJ...647..763M} {647, 763}

\bibitem[\protect\citeauthoryear{{Mayer}}{{Mayer}}{2013}]{mayer2013}
{Mayer} L.,  2013, \mn@doi [Classical and Quantum Gravity]
  {10.1088/0264-9381/30/24/244008}, \href
  {https://ui.adsabs.harvard.edu/abs/2013CQGra..30x4008M} {30, 244008}

\bibitem[\protect\citeauthoryear{{Mazzilli Ciraulo}, {Melchior}, {Maschmann},
  {Katkov}, {Halle}, {Combes}, {Gelfand}  \& {Al Yazeedi}}{{Mazzilli Ciraulo}
  et~al.}{2021}]{Mazzilli2021}
{Mazzilli Ciraulo} B.,  {Melchior} A.-L.,  {Maschmann} D.,  {Katkov} I.~Y.,
  {Halle} A.,  {Combes} F.,  {Gelfand} J.~D.,   {Al Yazeedi} A.,  2021, \mn@doi
  [\aap] {10.1051/0004-6361/202141319}, \href
  {https://ui.adsabs.harvard.edu/abs/2021A&A...653A..47M} {653, A47}

\bibitem[\protect\citeauthoryear{{Mihos} \& {Hernquist}}{{Mihos} \&
  {Hernquist}}{1994}]{Mihos1994}
{Mihos} J.~C.,  {Hernquist} L.,  1994, \mn@doi [\apj] {10.1086/174124}, \href
  {https://ui.adsabs.harvard.edu/abs/1994ApJ...427..112M} {427, 112}

\bibitem[\protect\citeauthoryear{{Moore}, {Katz}, {Lake}, {Dressler}  \&
  {Oemler}}{{Moore} et~al.}{1996}]{Moore1996}
{Moore} B.,  {Katz} N.,  {Lake} G.,  {Dressler} A.,   {Oemler} A.,  1996,
  \mn@doi [\nat] {10.1038/379613a0}, \href
  {https://ui.adsabs.harvard.edu/abs/1996Natur.379..613M} {379, 613}

\bibitem[\protect\citeauthoryear{{Naab}, {Jesseit}  \& {Burkert}}{{Naab}
  et~al.}{2006}]{Naab2006}
{Naab} T.,  {Jesseit} R.,   {Burkert} A.,  2006, \mn@doi [\mnras]
  {10.1111/j.1365-2966.2006.10902.x}, \href
  {https://ui.adsabs.harvard.edu/abs/2006MNRAS.372..839N} {372, 839}

\bibitem[\protect\citeauthoryear{{Navarro}, {Frenk}  \& {White}}{{Navarro}
  et~al.}{1996}]{Navarro1996}
{Navarro} J.~F.,  {Frenk} C.~S.,   {White} S. D.~M.,  1996, \mn@doi [\apj]
  {10.1086/177173}, \href
  {https://ui.adsabs.harvard.edu/abs/1996ApJ...462..563N} {462, 563}

\bibitem[\protect\citeauthoryear{{Ostriker} \& {Tremaine}}{{Ostriker} \&
  {Tremaine}}{1975}]{Ostriker1975}
{Ostriker} J.~P.,  {Tremaine} S.~D.,  1975, \mn@doi [\apjl] {10.1086/181992},
  \href {https://ui.adsabs.harvard.edu/abs/1975ApJ...202L.113O} {202, L113}

\bibitem[\protect\citeauthoryear{{Peschken}, {{\L}okas}  \&
  {Athanassoula}}{{Peschken} et~al.}{2020}]{Peschken2020}
{Peschken} N.,  {{\L}okas} E.~L.,   {Athanassoula} E.,  2020, \mn@doi [\mnras]
  {10.1093/mnras/staa299}, \href
  {https://ui.adsabs.harvard.edu/abs/2020MNRAS.493.1375P} {493, 1375}

\bibitem[\protect\citeauthoryear{{Pinna} et~al.,}{{Pinna}
  et~al.}{2019}]{Pinna2019}
{Pinna} F.,  et~al., 2019, \mn@doi [\aap] {10.1051/0004-6361/201833193}, \href
  {https://ui.adsabs.harvard.edu/abs/2019A&A...623A..19P} {623, A19}

\bibitem[\protect\citeauthoryear{{Pizzella}, {Corsini}, {Morelli}, {Sarzi},
  {Scarlata}, {Stiavelli}  \& {Bertola}}{{Pizzella}
  et~al.}{2002}]{Pizzella2002}
{Pizzella} A.,  {Corsini} E.~M.,  {Morelli} L.,  {Sarzi} M.,  {Scarlata} C.,
  {Stiavelli} M.,   {Bertola} F.,  2002, \mn@doi [\apj] {10.1086/340486}, \href
  {https://ui.adsabs.harvard.edu/abs/2002ApJ...573..131P} {573, 131}

\bibitem[\protect\citeauthoryear{{Poole}, {Mutch}, {Croton}  \&
  {Wyithe}}{{Poole} et~al.}{2017}]{Poole2017}
{Poole} G.~B.,  {Mutch} S.~J.,  {Croton} D.~J.,   {Wyithe} S.,  2017, \mn@doi
  [\mnras] {10.1093/mnras/stx2233}, \href
  {https://ui.adsabs.harvard.edu/abs/2017MNRAS.472.3659P} {472, 3659}

\bibitem[\protect\citeauthoryear{{Puech} et~al.,}{{Puech}
  et~al.}{2008}]{Puech2008}
{Puech} M.,  et~al., 2008, \mn@doi [\aap] {10.1051/0004-6361:20079313}, \href
  {https://ui.adsabs.harvard.edu/abs/2008A&A...484..173P} {484, 173}

\bibitem[\protect\citeauthoryear{{Pulsoni}, {Gerhard}, {Arnaboldi},
  {Pillepich}, {Nelson}, {Hernquist}  \& {Springel}}{{Pulsoni}
  et~al.}{2020}]{Pulsoni2020}
{Pulsoni} C.,  {Gerhard} O.,  {Arnaboldi} M.,  {Pillepich} A.,  {Nelson} D.,
  {Hernquist} L.,   {Springel} V.,  2020, \mn@doi [\aap]
  {10.1051/0004-6361/202038253}, \href
  {https://ui.adsabs.harvard.edu/abs/2020A&A...641A..60P} {641, A60}

\bibitem[\protect\citeauthoryear{{Purcell}, {Kazantzidis}  \&
  {Bullock}}{{Purcell} et~al.}{2009}]{Purcell2009}
{Purcell} C.~W.,  {Kazantzidis} S.,   {Bullock} J.~S.,  2009, \mn@doi [\apjl]
  {10.1088/0004-637X/694/2/L98}, \href
  {https://ui.adsabs.harvard.edu/abs/2009ApJ...694L..98P} {694, L98}

\bibitem[\protect\citeauthoryear{{Robertson}, {Cox}, {Hernquist}, {Franx},
  {Hopkins}, {Martini}  \& {Springel}}{{Robertson}
  et~al.}{2006}]{Robertson2006}
{Robertson} B.,  {Cox} T.~J.,  {Hernquist} L.,  {Franx} M.,  {Hopkins} P.~F.,
  {Martini} P.,   {Springel} V.,  2006, \mn@doi [\apj] {10.1086/500360}, \href
  {https://ui.adsabs.harvard.edu/abs/2006ApJ...641...21R} {641, 21}

\bibitem[\protect\citeauthoryear{{Sarzi}, {Ledo}  \& {Dotti}}{{Sarzi}
  et~al.}{2015}]{Sarzi2015}
{Sarzi} M.,  {Ledo} H.~R.,   {Dotti} M.,  2015, \mn@doi [\mnras]
  {10.1093/mnras/stv1497}, \href
  {https://ui.adsabs.harvard.edu/abs/2015MNRAS.453.1070S} {453, 1070}

\bibitem[\protect\citeauthoryear{{Sarzi} et~al.,}{{Sarzi}
  et~al.}{2016}]{Sarzi2016}
{Sarzi} M.,  et~al., 2016, \mn@doi [\mnras] {10.1093/mnras/stw099}, \href
  {https://ui.adsabs.harvard.edu/abs/2016MNRAS.457.1804S} {457, 1804}

\bibitem[\protect\citeauthoryear{{Schweizer}}{{Schweizer}}{1996}]{Schweizer1996}
{Schweizer} F.,  1996, \mn@doi [\aj] {10.1086/117765}, \href
  {https://ui.adsabs.harvard.edu/abs/1996AJ....111..109S} {111, 109}

\bibitem[\protect\citeauthoryear{{Schweizer} \& {Seitzer}}{{Schweizer} \&
  {Seitzer}}{1992}]{Schweizer1992}
{Schweizer} F.,  {Seitzer} P.,  1992, \mn@doi [\aj] {10.1086/116296}, \href
  {https://ui.adsabs.harvard.edu/abs/1992AJ....104.1039S} {104, 1039}

\bibitem[\protect\citeauthoryear{{Serra} et~al.,}{{Serra}
  et~al.}{2016}]{Serra2017}
{Serra} P.,  et~al., 2016, in MeerKAT Science: On the Pathway to the SKA. p.~8
  (\mn@eprint {arXiv} {1709.01289}), \mn@doi{10.22323/1.277.0008}

\bibitem[\protect\citeauthoryear{{Shah} et~al.,}{{Shah}
  et~al.}{2022}]{Shah2022}
{Shah} E.~A.,  et~al., 2022, \mn@doi [\apj] {10.3847/1538-4357/ac96eb}, \href
  {https://ui.adsabs.harvard.edu/abs/2022ApJ...940....4S} {940, 4}

\bibitem[\protect\citeauthoryear{{Shen}, {Mo}, {White}, {Blanton}, {Kauffmann},
  {Voges}, {Brinkmann}  \& {Csabai}}{{Shen} et~al.}{2003}]{Shen2003}
{Shen} S.,  {Mo} H.~J.,  {White} S. D.~M.,  {Blanton} M.~R.,  {Kauffmann} G.,
  {Voges} W.,  {Brinkmann} J.,   {Csabai} I.,  2003, \mn@doi [\mnras]
  {10.1046/j.1365-8711.2003.06740.x}, \href
  {https://ui.adsabs.harvard.edu/abs/2003MNRAS.343..978S} {343, 978}

\bibitem[\protect\citeauthoryear{{Sotillo-Ramos} et~al.,}{{Sotillo-Ramos}
  et~al.}{2022}]{Sotillo2022}
{Sotillo-Ramos} D.,  et~al., 2022, \mn@doi [\mnras] {10.1093/mnras/stac2586},
  \href {https://ui.adsabs.harvard.edu/abs/2022MNRAS.516.5404S} {516, 5404}

\bibitem[\protect\citeauthoryear{{Sparre} \& {Springel}}{{Sparre} \&
  {Springel}}{2017}]{Sparre2017}
{Sparre} M.,  {Springel} V.,  2017, \mn@doi [\mnras] {10.1093/mnras/stx1516},
  \href {https://ui.adsabs.harvard.edu/abs/2017MNRAS.470.3946S} {470, 3946}

\bibitem[\protect\citeauthoryear{{Springel}, {Di Matteo}  \&
  {Hernquist}}{{Springel} et~al.}{2005}]{Springel2005}
{Springel} V.,  {Di Matteo} T.,   {Hernquist} L.,  2005, \mn@doi [\apjl]
  {10.1086/428772}, \href
  {https://ui.adsabs.harvard.edu/abs/2005ApJ...620L..79S} {620, L79}

\bibitem[\protect\citeauthoryear{{Stewart}, {Bullock}, {Wechsler}, {Maller}  \&
  {Zentner}}{{Stewart} et~al.}{2008}]{Stewart2008}
{Stewart} K.~R.,  {Bullock} J.~S.,  {Wechsler} R.~H.,  {Maller} A.~H.,
  {Zentner} A.~R.,  2008, \mn@doi [\apj] {10.1086/588579}, \href
  {https://ui.adsabs.harvard.edu/abs/2008ApJ...683..597S} {683, 597}

\bibitem[\protect\citeauthoryear{{Tonnesen} \& {Cen}}{{Tonnesen} \&
  {Cen}}{2012}]{Tonnesen2012}
{Tonnesen} S.,  {Cen} R.,  2012, \mn@doi [\mnras]
  {10.1111/j.1365-2966.2012.21637.x}, \href
  {https://ui.adsabs.harvard.edu/abs/2012MNRAS.425.2313T} {425, 2313}

\bibitem[\protect\citeauthoryear{{Toomre}}{{Toomre}}{1977}]{Toomre1977}
{Toomre} A.,  1977, in {Tinsley} B.~M.,  {Larson} Richard B.~Gehret D.~C.,
  eds, Evolution of Galaxies and Stellar Populations. p.~401

\bibitem[\protect\citeauthoryear{{Toomre} \& {Toomre}}{{Toomre} \&
  {Toomre}}{1972}]{Toomre1972}
{Toomre} A.,  {Toomre} J.,  1972, \mn@doi [\apj] {10.1086/151823}, \href
  {https://ui.adsabs.harvard.edu/abs/1972ApJ...178..623T} {178, 623}

\bibitem[\protect\citeauthoryear{{Van Wassenhove}, {Capelo}, {Volonteri},
  {Dotti}, {Bellovary}, {Mayer}  \& {Governato}}{{Van Wassenhove}
  et~al.}{2014}]{VanWassenhove2014}
{Van Wassenhove} S.,  {Capelo} P.~R.,  {Volonteri} M.,  {Dotti} M.,
  {Bellovary} J.~M.,  {Mayer} L.,   {Governato} F.,  2014, \mn@doi [\mnras]
  {10.1093/mnras/stu024}, \href
  {https://ui.adsabs.harvard.edu/abs/2014MNRAS.439..474V} {439, 474}

\bibitem[\protect\citeauthoryear{{Vasiliev}}{{Vasiliev}}{2019}]{Vasiliev2019}
{Vasiliev} E.,  2019, \mn@doi [\mnras] {10.1093/mnras/sty2672}, \href
  {https://ui.adsabs.harvard.edu/abs/2019MNRAS.482.1525V} {482, 1525}

\bibitem[\protect\citeauthoryear{{White} \& {Rees}}{{White} \&
  {Rees}}{1978}]{White1978}
{White} S.~D.~M.,  {Rees} M.~J.,  1978, \mn@doi [\mnras]
  {10.1093/mnras/183.3.341}, \href
  {https://ui.adsabs.harvard.edu/abs/1978MNRAS.183..341W} {183, 341}

\bibitem[\protect\citeauthoryear{{Woods} \& {Geller}}{{Woods} \&
  {Geller}}{2007}]{Woods2007}
{Woods} D.~F.,  {Geller} M.~J.,  2007, \mn@doi [\aj] {10.1086/519381}, \href
  {https://ui.adsabs.harvard.edu/abs/2007AJ....134..527W} {134, 527}

\bibitem[\protect\citeauthoryear{{Yadav} \& {Chen}}{{Yadav} \&
  {Chen}}{2018}]{Yadav2018}
{Yadav} J.~K.,  {Chen} X.,  2018, \mn@doi [Journal of Astrophysics and
  Astronomy] {10.1007/s12036-018-9521-x}, \href
  {https://ui.adsabs.harvard.edu/abs/2018JApA...39...31Y} {39, 31}

\bibitem[\protect\citeauthoryear{{Yun} et~al.,}{{Yun} et~al.}{2019}]{Yun2019}
{Yun} K.,  et~al., 2019, \mn@doi [\mnras] {10.1093/mnras/sty3156}, \href
  {https://ui.adsabs.harvard.edu/abs/2019MNRAS.483.1042Y} {483, 1042}

\bibitem[\protect\citeauthoryear{{de Vaucouleurs}}{{de
  Vaucouleurs}}{1948}]{Vaucouleurs1948}
{de Vaucouleurs} G.,  1948, Annales d'Astrophysique, \href
  {https://ui.adsabs.harvard.edu/abs/1948AnAp...11..247D} {11, 247}

\bibitem[\protect\citeauthoryear{van~den Bosch, {Ferrarese}, {Jaffe}, {Ford}
  \& {O'Connell}}{van~den Bosch et~al.}{1994}]{Bosch1994}
van~den Bosch F.~C.,  {Ferrarese} L.,  {Jaffe} W.,  {Ford} H.~C.,   {O'Connell}
  R.~W.,  1994, \mn@doi [\aj] {10.1086/117179}, \href
  {https://ui.adsabs.harvard.edu/abs/1994AJ....108.1579V} {108, 1579}

\makeatother
\end{thebibliography}
\normalsize








\bsp	
\label{lastpage}
\end{document}